\documentclass{osa-article}
\journal{osac}

\usepackage{hyperref}
\usepackage{amsmath,calligra,amssymb,mathtools}
\usepackage{color}

\newcommand{\veps}{\varepsilon}
\newcommand{\ld}{\lambda}
\newcommand{\Om}{\Omega}
\newcommand{\om}{\omega}
\newcommand{\sg}{\sigma}
\newcommand{\vr}{{\bf r}}
\newcommand{\vF}{{\bf F}}
\newcommand{\vE}{{\bf E}}
\newcommand{\ve}{{\bf e}}
\newcommand{\vY}{{\bf Y}}
\newcommand{\vH}{{\bf H}}
\newcommand{\Gm}{\Gamma}
\newcommand{\gm}{\gamma}

\begin{document}

\title{STRATIFY: a comprehensive and versatile MATLAB code for a multilayered sphere}

\author{
Ilia L. Rasskazov,\authormark{1,*}
P. Scott Carney,\authormark{1} and 
Alexander Moroz\authormark{2}}
\address{
\authormark{1} The Institute of Optics, University of Rochester, Rochester, NY 14627, USA\\
\authormark{2} Wave-scattering.com (e-mail: wavescattering@yahoo.com)
}

\email{\authormark{*}irasskaz@ur.rochester.edu}

\begin{abstract}
We present a computer code for calculating near- and far-field electromagnetic properties of multilayered spheres. 
STRATIFY is one-of-a-kind open-source package which allows for the efficient calculation of electromagnetic near-field, energy density, total electromagnetic energy, radiative and non-radiative decay rates of a dipole emitter located in any (non-absorbing) shell (including a host medium), and fundamental cross-sections of a multilayered sphere, all within a single program.
The developed software is typically more than $50$ times faster than freely available packages based on boundary-element-method. 
Because of its speed and broad applicability, our package is a valuable tool for analysis of numerous light scattering problems, including, but not limited to fluorescence enhancement, upconversion, downconversion, second harmonic generation, surface enhanced Raman spectroscopy.
The software is available for download from GitLab \url{https://gitlab.com/iliarasskazov/stratify}.
\end{abstract}

\section{Introduction}
Multilayered spherical nanoparticles are fundamental building blocks for many vital applications in physics and chemistry:
tailored scattering ~\cite{Neeves1988,Neeves1989,Zhou1994,Averitt1997,Oldenburg1998,Graf2002,Hasegawa2006} and nonlinear optics~\cite{Neeves1988,Neeves1989,Pu2010,Scherbak2018},
photonic crystals~\cite{Moroz1999,Zhang2000,Moroz2000,Velikov2002,Moroz2002},
sensing~\cite{Raschke2004,Jain2007,Ochsenkuhn2009},
photothermal cancer treatment~\cite{Hirsch2003,Ayala-Orozco2014,Zakomirnyi17JQSRT,Kostyukov19JQSRT},
solar energy harvesting~\cite{Phan2018a,Wang2018}, 
photovoltaics~\cite{Xu2020},
fluorescence~\cite{Chew1976,Tovmachenko2006,Zhang2006,Ayala-Orozco2014a,Sakamoto2017,Sun2020}, photoluminescence~\cite{Naiki2017} and
upconversion~\cite{Zhang2010a,Priyam2012,Yuan2012a,Kannan2013,Ding2014,Xu2014,Qin2016,Wang2016a,Rasskazov18OMEx} enhancement,
surface-enhanced Raman scattering (SERS)~\cite{Chew1976,Lim2011,Li2017a},
surface plasmon amplification by stimulated emission of radiation (spaser)~\cite{Noginov2009,Calander2012,Baranov2013,Arnold2016,Passarelli2016,Galanzha2017}, and cloaking~\cite{Alu2008,Monticone2013,Sheverdin2019,Tsakmakidis2019}.
To date, such matryoshkas with various combinations of dielectric and metal shells have well-established protocols for the efficient and controllable synthesis~\cite{Graf2002,Hirsch2006,Jankiewicz2012,Montano-Priede2017,Montano-Priede2017a,Wang2018h}, greatly improved on initial synthesis attempts~\cite{Zhou1994,Averitt1997,Oldenburg1998}.
Theoretical and numerical studies in these fields are inevitable for a thoughtful design of experiment and reliable interpretation of measured data.
The theory of electromagnetic light scattering from multilayered spheres has a long history since the pioneering work of Aden and Kerker for two concentric spheres~\cite{Aden1951}.
An impressive number of closed-form solutions and discussions on coated~\cite{Kaiser1994,Lock1994} and general multilayered~\cite{Sinzig1994} spheres have been reported.
There is a thorough understanding of scattering~\cite{Bhandari1985} and absorption~\cite{Mackowski1990} of light from multilayered spheres, including aspects for various types of illuminations~\cite{Li2007,Wang2011,Onofri1995,Wu1997a},
spontaneous decay rates of a dipole emitter in a presence of a multilayered sphere~\cite{Moroz2005,Moroz2005a},
electromagnetic energy~\cite{Rasskazov19JOSAA} and near-field~\cite{Schelm2005a} distribution within and near the multilayered spheres,
and the respective strategies for numerical implementation of the developed  theories~\cite{Toon1981,Wu1991,Yang2003,Majic2020}.

For a single shell, a number of features can be qualitatively understood from the quasi-static analysis~\cite{Neeves1989}, a full, quantitative understanding of the mechanisms underlying the applications of multilayered spheres is usually quite involved, requiring sophisticated and complex theoretical and numerical studies.
The latter are often handled with commercial brute-force finite-difference time-domain (FDTD), finite-element-method (FEM) or boundary-element-method (BEM) solvers, which provide accurate results at relatively high computational price.
However, spherical multilayered particles are computationally feasible \textit{per se} due to the existence of the closed-form analytical solutions inherently convenient for computational implementation.
Nonetheless, there are few \textit{freely available}, \textit{user-friendly} and \textit{comprehensive} codes for light scattering from a general multilayered spheres based on these solutions~\cite{wvsccodes}.
In most of the cases, a very limited number of properties are calculated within a single freely available package.
Fundamental cross-sections (i.e. scattering, absorption and extinction) and/or near fields are the usual choice implemented in freely available codes~\cite{Bohren1998,Ladutenko2017}, and in a number of so-called ``online Mie calculators'', which commonly handle only homogeneous spheres with a rare exceptions for core-shells and even more rarely for general multilayered spheres.
Computation of the orientation-averaged electric or magnetic field intensities (i.e. averaged over a spherical surface of a given fixed radius $r$), spontaneous decay rates and other quantities are almost unavailable, yet generally very useful.
Here we present free MATLAB code based on compact and easy-to-implement transfer-matrix formalism, which can be conveniently used to handle most of the problems of light scattering from multilayered spheres by using appropriate combinations and manipulations of transfer-matrices.
The use of closed-form solutions significantly enhances the performance of the code compared with brute-force solvers without sacrificing the accuracy.

Our recursive transfer-matrix method (RTMM) has been inspired by the success of such a RTMM for {\em planar} stratified media developed by Abel\`{e}s \cite{Abeles1948,Abeles1950,Abeles1950a}, summarized in Born \& Wolf classical textbook~\cite[Sec.1.6]{Born2013}, and can be seen as its analogue for {\em spherical} interfaces. 
From a historical perspective, the RTMM presented here had been developed as early as in 1998 and implemented in the F77 \verb| sphere.f| code~\cite{wvsccodes}. 
It was first applied to a number of different theoretical settings with external plane wave source~\cite{Moroz1999,Moroz2000,Moroz2002}, and successfully tested against experiment~\cite{Velikov2002,Graf2002}.
Later it was extended to incorporate a dipole source~\cite{Moroz2005,Moroz2005a}, energy calculations~\cite{Rasskazov19JOSAA}, and has enabled exhaustive optimization of plasmon-enhanced fluorescence~\cite{Sun2020}.

The paper is organized as follows. 
In Sec.~\ref{sec:theory}, we provide an overview of the RTMM for multilayered spheres and formulate the fundamental properties of spheres.
In Sec.~\ref{sec:code}, we describe the developed code and benchmark it with robust BEM~\cite{GarciadeAbajo2002} implemented in freely available ``MNPBEM'' MATLAB package~\cite{Hohenester2012,Hohenester2014,Waxenegger2015,Hohenester2018}.
In Sec.~\ref{sec:disc}, we discuss possible application of the code in SERS, plasmon-enhanced fluorescence, upconversion, and end with conclusive remarks and propose further developments of the package.
Gaussian units are used throughout the paper.

\section{Theory}
\label{sec:theory}

\subsection{Recursive Transfer-Matrix method}
\label{sec:tmatrix}
\begin{figure}[b]
    \centering
    \includegraphics[width=4in]{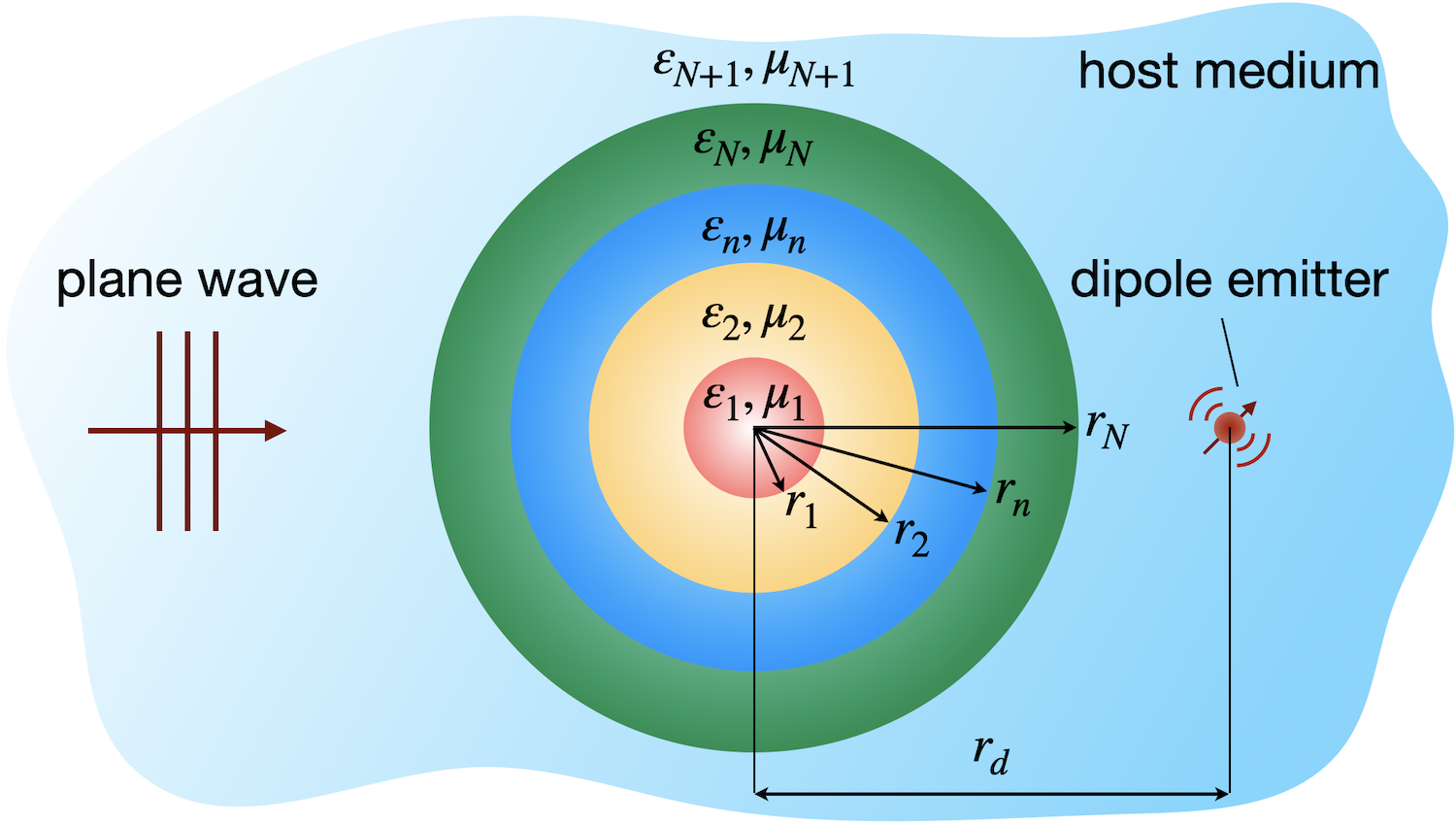}
    \caption{
    A sketch of the multilayered sphere in a homogeneous isotropic host medium with permittivity $\veps_h=\veps_{N+1}$ and permeability $\mu_h=\mu_{N+1}$.
    Center of the sphere is located at the origin of the reference system.}
    \label{fig:scheme}
\end{figure}

Consider a multilayered sphere with $N$ concentric shells as shown in Fig.~\ref{fig:scheme}.
The sphere core counts as a shell with number $n=1$ and the host medium is the $n=N+1$ shell.
Occasionally, the host medium will be denoted as $n=h$.
Each shell has the outer radius $r_n$ and is assumed to be homogeneous and isotropic with scalar permittivity $\veps_n$ and permeability $\mu_n$.
Respective refractive indices are $\eta_n = \sqrt{\veps_n \mu_n}$.
We assume that the multilayered sphere is illuminated with a harmonic electromagnetic wave (either a plane wave or a dipole source) having vacuum wavelength $\ld$.
Corresponding wave vector in the $n$-th shell is $k_n = \eta_n\om/c = 2\pi\eta_n/\ld$, where $c$ is the speed of light in vacuum, and $\om$ is frequency. 
Electromagnetic fields in any shell are described by the stationary macroscopic Maxwell's equations (with time dependence $e^{-i\om t}$ assumed and suppressed throughout the paper):

\begin{equation}
\vE=\frac{ic}{\om\veps}\,
         \left(\mbox{\boldmath $\nabla$}\times \vH\right) \ , \qquad
\vH= -\frac{i c}{\om\mu}\,\left(\mbox{\boldmath $\nabla$}\times\vE\right) \ ,
\label{mex}
\end{equation}
where the permittivity $\veps$ and permeability $\mu$ are scalars.
Following the notation of \cite{Moroz2005}, the basis of normalized (the normalization here refers to angular integration) {\em transverse} vector multipole fields $\mbox{\boldmath $\nabla$}\cdot\vF_{pL}\equiv 0$ that satisfy the vector Helmholtz equation

\begin{equation}
\mbox{\boldmath $\nabla$}\times \left[\mbox{\boldmath $\nabla$} \times \vF_{pL}(k,\vr) \right] = k^2 \vF_{pL}(k,\vr) 
\end{equation}
for $n$-th shell, $1\leq n\leq N+1$, can be formed as:

\begin{equation}
\begin{split}
\vF&_{ML}(k_n,\vr) = f_{M L} (k_n r) \vY^{(m)}_L(\vr) \ , \\
\vF&_{EL}(k_n,\vr) = \frac{1}{k_n r}\left\{\sqrt{\ell(\ell+1)}f_{E L} (k_n r)\vY^{(o)}_L(\vr) + \dfrac{\rm d}{{\rm d}r} \left[r f_{E L}(k_n r)\right]\vY^{(e)}_L(\vr)\right\} \ ,
\end{split}
\label{fvmultip}
\end{equation}
where
\begin{equation}
\tilde\vF_{EL}(k_n,\vr) = \frac{1}{k_n}\mbox{\boldmath $\nabla$}
     \times \vF_{ML}(k_n,\vr) \ , \quad
\tilde\vF_{ML}(k_n,\vr) = \frac{1}{k_n}\mbox{\boldmath $\nabla$}
    \times \vF_{EL}(k_n,\vr) \ ,
\label{fvmultipr}
\end{equation}
with $\tilde{f}_{EL}=f_{ML}$ and $\tilde{f}_{ML}=f_{EL}$. 
Here $L=\ell,m$ is a composite angular momentum index and $f_{pL}$ is a suitable linear combination of spherical Bessel functions.
Provided that the multipole fields in \eqref{fvmultip} represent ${\bf E}$, the respective subscripts $p=M$ and $p=E$ denote the {\em magnetic}, or {\em transverse electric} (TE), and {\em electric}, or {\em transverse magnetic} (TM), polarizations~\cite{Jackson1999}.
The so-called magnetic, $\vY^{(m)}_L$, longitudinal, $\vY^{(o)}_L$, and electric, $\vY^{(e)}_L$, vector spherical harmonics of degree $\ell$ and order $m$, are defined in \textit{spherical} $(\varphi, \vartheta, r)$ coordinates as~\cite{Kerker1980,Bohren1998,Jackson1999,Mishchenko2002}
\begin{equation}
\begin{split}
\vY^{(m)}_L &= i \sqrt{ \frac{(\ell-m)!}{(\ell+m)!} } \sqrt{\frac{2\ell+1}{4\pi \ell(\ell+1)}} \left[ 
    \hat\ve_\vartheta \frac{im P_\ell^m (\cos\vartheta)}{\sin\vartheta} 
    - \hat\ve_\varphi \frac{{\rm d} P_\ell^m  (\cos\vartheta)}{{\rm d}\vartheta} 
    \right] \exp\left(im\varphi\right) \ ,
\\
\vY^{(o)}_L &= i \sqrt{ \frac{(\ell-m)!}{(\ell+m)!} } \sqrt{\frac{2\ell+1}{4\pi }} P_\ell^m (\cos\vartheta) \exp\left(im\varphi\right) \hat\ve_r \ ,
\\
\vY^{(e)}_L &= i \sqrt{ \frac{(\ell-m)!}{(\ell+m)!} } \sqrt{\frac{2\ell+1}{4\pi \ell(\ell+1)}} \left[ 
    \hat\ve_\vartheta \frac{{\rm d} P_\ell^m (\cos\vartheta)}{{\rm d}\vartheta} 
    + \hat\ve_\varphi \frac{im P_\ell^m (\cos\vartheta)}{\sin\vartheta} 
    \right] \exp\left(im\varphi\right) \ ,
\label{eq:Ymoe}
\end{split}
\end{equation}
where $\hat\ve_\varphi$, $\hat\ve_\vartheta$, and $\hat\ve_r$ are corresponding unit vectors, and $P^m_\ell(x)$ are the associated Legendre functions of the first kind~\cite{Abramowitz1973} of degree $\ell$ and order $m$:

\begin{equation}
P^m_\ell(x)=\frac{(-1)^m}{2^\ell \ell!} (1-x^2)^{m/2} \frac{{\rm d}^{\ell+m}}{{\rm d}x^{\ell+m}} (x^2-1)^\ell \ .
\nonumber 
\end{equation}
%

General solution for the electric field in the $n$-th shell, $1\leq n\leq N+1$, is~\cite[Eq. (8)]{Moroz2005}:
\begin{equation}
\vE_p (\vr) 
= \sum_L \vF_{pL}(k_n, \vr)
= \sum_{L} \left[ A_{pL} (n)\, {\bf J}_{pL}(k_n, \vr) + B_{pL}(n)\, \vH_{pL}(k_n, \vr)\right] \ .
\label{eq:Efld}
\end{equation}
In order to emphasize and for the sake of tracking the Bessel function dependence, the multipoles $\vF_{pL}$ in Eq.~\eqref{eq:Efld} are denoted as ${\bf J}_{pL}$ and $\vH_{pL}$ for the respective cases that $f_{p\ell}=j_\ell$ and $f_{p\ell}=h_\ell^{(1)}$, where $j_\ell$ and $h_\ell^{(1)}$ being the spherical Bessel functions of the first and third kind, correspondingly.

Given the general solution \eqref{eq:Efld}, the corresponding expansion of magnetic field $\vH$ 
follows from that of the electric field $\vE$ by the stationary macroscopic Maxwell's equations (\ref{mex}) on using relations (\ref{fvmultipr}) \cite[Eqs. (9)-(10)]{Moroz2005}: 

\begin{equation}
\begin{split}
\vH_E (\vr) 
& = -i \dfrac{\veps_n}{\mu_n} \sum_{L} \left[ A_{EL}(n)\, {\bf J}_{ML}(k_n, \vr) 
 + B_{EL}(n)\, \vH_{ML}(k_n, \vr)\right] \ , \\
\vH_M (\vr) 
& = -i \dfrac{\veps_n}{\mu_n} \sum_{L} \left[ A_{ML}(n)\, {\bf J}_{EL}(k_n, \vr) 
  + B_{ML}(n)\, \vH_{EL}(k_n, \vr)\right] \ .
\end{split}
\label{eq:Hfld}
\end{equation}
The $\vH_{pL}$'s in Eqs.~\eqref{eq:Efld}-\eqref{eq:Hfld} with angular-momentum index $L$ refer
to a basis multipole which distinguishes them from the magnetic field $\vH_{p}$.

Very much as in the case of planar stratified media~\cite{Abeles1948,Abeles1950,Abeles1950a,Born2013},
our RTMM for spherical interfaces exploits to the maximum the property that once the coefficients 
$A_{pL}(n+1)$ and $B_{pL}(n+1)$ on one side of a $n$-th shell interface are known, the coefficients $A_{pL}(n)$ and $B_{pL}(n)$ on the other side of the shell interface can be unambiguously determined, and vice versa. Schematically,

\begin{equation}
\begin{pmatrix} A_{pL}(n+1)\\ B_{pL}(n+1) \end{pmatrix} = T_{p\ell}^+(n) \begin{pmatrix} A_{pL}(n)\\B_{pL}(n) \end{pmatrix} \ , \qquad 
\begin{pmatrix} A_{p L}(n)\\ B_{p L}(n) \end{pmatrix} = T_{p\ell}^-(n) \begin{pmatrix} A_{p L}(n+1)\\B_{p L}(n+1) \end{pmatrix} \ ,
\label{eq:trm}
\end{equation}
where $T_{p\ell}^\pm (n)$ are $2\times 2$ {\em transfer matrices}~\cite{Moroz2005}. 
The matrices, which are the fundamental quantities of our approach, can be viewed as a 
kind of {\em raising} and {\em lowering ladder operators} of quantum mechanics. In the present case, they lower and raise the argument $n$ of expansion coefficients $A_{pL}(n)$ and $B_{pL}(n)$.
The transfer matrices are interrelated by 
\begin{equation}
\left[ T_{p\ell}^+(n)\right]^{-1} = T_{p\ell}^-(n) \ ,
\qquad
\left[ T_{p\ell}^-(n)\right]^{-1} = T_{p\ell}^+(n) \ ,
\label{tinv}
\end{equation}
and are {\em unambiguously} determined by matching fields across the $n$-th shell interface, i.e. 
requiring that the {\em tangential} components of $\vE$ and $\vH$ are continuous.  
One finds~\cite{Moroz2005}:

\begin{equation}
T^-_{M\ell}(n) =
- i \begin{pmatrix}
  \tilde{\eta} \zeta_\ell'(x) \psi_\ell(\tilde{x}) - \tilde{\mu} \zeta_\ell(x)\psi_\ell'(\tilde{x}) 
& \tilde{\eta} \zeta_\ell'(x) \zeta_\ell(\tilde{x}) - \tilde{\mu} \zeta_\ell(x) \zeta_\ell'(\tilde{x}) \\
- \tilde{\eta} \psi_\ell'(x)\psi_\ell(\tilde{x}) + \tilde{\mu} \psi_\ell(x)\psi_\ell'(\tilde{x})
& - \tilde{\eta} \psi_\ell'(x) \zeta_\ell(\tilde{x}) + \tilde{\mu} \psi_\ell(x)\zeta_\ell'(\tilde{x})
\end{pmatrix} \ ,
\label{eq:bacmtm}
\end{equation}

\begin{equation}
T^-_{E\ell}(n) = 
- i \begin{pmatrix}
  \tilde{\mu} \zeta_\ell'(x)\psi_\ell(\tilde{x}) - \tilde{\eta} \zeta_\ell(x)\psi_\ell'(\tilde{x}) 
& \tilde{\mu} \zeta_\ell'(x) \zeta_\ell(\tilde{x}) - \tilde{\eta} \zeta_\ell(x) \zeta_\ell'(\tilde{x}) \\
- \tilde{\mu} \psi_\ell'(x)\psi_\ell(\tilde{x}) + \tilde{\eta} \psi_\ell(x)\psi_\ell'(\tilde{x})
& - \tilde{\mu} \psi_\ell'(x) \zeta_\ell(\tilde{x}) + \tilde{\eta} \psi_\ell(x)\zeta_\ell'(\tilde{x})
\end{pmatrix} \ ,
\label{eq:bacetm}
\end{equation}

\begin{equation}
T^+_{M\ell}(n) = 
- i \begin{pmatrix}
  \zeta_\ell'(\tilde{x}) \psi_\ell(x)/\tilde{\eta} - \zeta_\ell(\tilde{x})\psi_\ell'(x)/\tilde{\mu}
& \zeta_\ell'(\tilde{x}) \zeta_\ell(x)/\tilde{\eta} - \zeta_\ell(\tilde{x})\zeta_\ell'(x)/\tilde{\mu} \\
- \psi_\ell'(\tilde{x})\psi_\ell(x)/\tilde{\eta} + \psi_\ell(\tilde{x})\psi_\ell'(x) /\tilde{\mu}
& - \psi_\ell'(\tilde{x}) \zeta_\ell(x)/\tilde{\eta} + \psi_\ell(\tilde{x})\zeta_\ell'(x)/\tilde{\mu}
\end{pmatrix} \ ,
\label{eq:formtm}
\end{equation}

\begin{equation}
T^+_{E\ell}(n) =
- i \begin{pmatrix}
  \zeta_\ell'(\tilde{x})\psi_\ell(x)/\tilde{\mu}  - \zeta_\ell(\tilde{x})\psi_\ell'(x) /\tilde{\eta}
& \zeta_\ell'(\tilde{x}) \zeta_\ell(x)/\tilde{\mu} - \zeta_\ell(\tilde{x}) \zeta_\ell'(x) /\tilde{\eta} \\
- \psi_\ell'(\tilde{x})\psi_\ell(x)/\tilde{\mu} + \psi_\ell(\tilde{x})\psi_\ell'(x) /\tilde{\eta}
& - \psi_\ell'(\tilde{x}) \zeta_\ell(x)/\tilde{\mu} + \psi_\ell(\tilde{x})\zeta_\ell'(x) /\tilde{\eta}
\end{pmatrix} \ ,
\label{eq:foretm}
\end{equation}
\noindent
where $\psi_\ell(x) = xj_\ell(x)$ and $\zeta_\ell(x) = x h^{(1)}_\ell(x)$ are the Riccati-Bessel functions, prime denotes the derivative with respect to the argument in parentheses, 
$x_n=k_n r_n$ and $\tilde{x}_n=x_n / \tilde{\eta}_n=k_{n+1} r_n$ are the internal and external dimensionless size parameters, $\tilde{\eta}_n=\eta_n/\eta_{n+1}$ and $\tilde{\mu}_n = \mu_n/\mu_{n+1}$ are relative refractive indices and permeabilities.
For the sake of clarity, the $n$-subscript has been suppressed on the rhs of Eqs. \eqref{eq:bacmtm}~--~\eqref{eq:foretm}. 
The above expressions for $T^{\pm}_{p\ell}(n)$ are general, independent on the incident field, and valid for any homogeneous and isotropic medium, including \textit{gain} ($\Im(\veps_n)<0$) or {\em magnetic} materials ($\mu_n \neq 1$).

The formalism becomes compact upon the use of the \textit{composite} transfer matrices ${\cal T}_{p\ell}(n)$ and ${\cal M}_{p\ell}(n)$ defined as ordered (from the left to the right) products of the constituent {\em raising} and {\em lowering} $2\times 2$ matrices from Eqs.~\eqref{eq:bacmtm}-\eqref{eq:foretm}:

\begin{equation}
{\cal T}_{p\ell}(n) = \prod_{j=n-1}^{1} T_{p\ell}^+(j) \ , \qquad 
{\cal M}_{p\ell}(n) = \prod_{j=n}^N T_{p\ell}^-(j) \ .
\label{eq:ordprod}
\end{equation}
Composite matrices ${\cal T}_{p\ell}(n)$ and ${\cal M}_{p\ell} (n)$ transfer expansion coefficients to the $n$-th shell from the sphere core or from the surrounding medium, respectively. 
Note that ${\cal T}_{p\ell}(n)$ are defined for $2 \leq n\leq N+1$, while ${\cal M}_{p\ell}(n)$ are defined for $1\leq n\leq N$. They are used in our formalism to chiefly relate the expansion coefficients in the core to those in a source region. For example, ${\cal T}_{p\ell}(N+1)$ enables one to obviate all the intermediary shell interfaces 
and to relate the expansion coefficients in the core directly to those in the surrounding host medium,
\begin{equation}
\begin{pmatrix} A_{pL}(N+1)\\ B_{pL}(N+1) \end{pmatrix} = {\cal T}_{p\ell}(N+1) 
\begin{pmatrix} A_{pL}(1)\\B_{pL}(1) \end{pmatrix} \ .
\label{eq:trm1toN}
\end{equation}
The latter formally reduces the problem of a {\em multilayered} sphere with a source in the host medium
to that of a {\em homogeneous} sphere. Analogously to \eqref{tinv}:

\begin{equation}
\left[{\cal T}_{p\ell}(N+1)\right]^{-1} = {\cal M}_{p\ell}(1) \ ,
\qquad
\left[{\cal M}_{p\ell}(1)\right]^{-1} = {\cal T}_{p\ell}(N+1) \ .
\label{tminv}
\end{equation}
The reader familiar with the RTMM for planar stratified media (e.g. multilayer coatings and interference filters) will recognize in the relations \eqref{eq:trm}, \eqref{tinv}, \eqref{eq:ordprod}, \eqref{eq:trm1toN}, \eqref{tminv} familiar properties of the constituent and composite transfer matrices for planar interfaces~\cite{Abeles1948,Abeles1950,Abeles1950a,Born2013}.

In order to {\em unambiguously} determine the expansion coefficients $A_{pL}(n)$ and $B_{pL}(n)$ in each shell,
one has to impose {\em two} boundary conditions, which is the subject of the following section.

\subsection{Boundary conditions}
The total number of different sets of expansion coefficients comprising all $j$'s from the interval
$1\leq j\leq N+1$ is larger by two than the number of corresponding equations. 
Therefore, boundary conditions have to be imposed to unambiguously determine the 
expansion coefficients at any shell. They are


\begin{enumerate}

\item The {\em regularity} condition of the solution at the sphere origin, which eliminates $h_{\ell}^{(1)}(0)\to \infty$ for $f_{p\ell}$ in Eq.~\eqref{fvmultip}: 

\begin{equation}
B_{EL}(1)=B_{ML}(1)\equiv 0 \ .
\label{dipregbc}
\end{equation}

\item For a source located \textit{outside} a sphere,
the $A_{pL}(N+1)$ coefficients at any given frequency $\om$ are equal to the expansion coefficients of an incident electromagnetic field in spherical coordinates.

\end{enumerate}
The {\em regularity} condition (\ref{dipregbc}) alone suffices to 
unambiguously determine the $m$-independent (due to spherical symmetry of a problem) ratio~\cite{Jackson1999}:

\begin{equation}
{\cal T}_{21;p\ell}(N+1)/{\cal T}_{11;p\ell}(N+1) = \mathbb{T}_{p\ell},
\label{dcrat} 
\end{equation}
which defines the familiar T-matrix of scattering theory, $\mathbb{T}_{p\ell}$, where $i,j$ in ${\cal T}_{ij;p\ell}(n)$ label
the $(i,j)$-th element of the $2\times 2$ matrix ${\cal T}_{p\ell}(n)$. 
Note that $\mathbb{T}_{p\ell}$ are nothing but familiar expansion coefficients $a_\ell$ and $b_\ell$ in Bohren and Huffman's representation~\cite[Eqs. (4.56),(4.57)]{Bohren1998}, but with the opposite sign: $\mathbb{T}_{E\ell} = - a_{\ell}$ and $\mathbb{T}_{M\ell} = - b_{\ell}$.

Corresponding closed-form analytic expressions for applying the second boundary condition
are presented in (i) Refs.~\cite{Mishchenko2002},\cite[Eq. (10)]{Rasskazov19JOSAA} for the plane electromagnetic wave, and in (ii) Ref.~\cite[Eqs. (44),(50)]{Moroz2005} for the electric dipole source. 
In the case of an elementary dipole radiating \textit{inside} a sphere, there is no source outside a sphere, and the second boundary condition reduces to $A_{pL}(N+1)=0$~\cite{Moroz2005}.

\subsection{Far-field properties}
\label{sec:farfld}
Fundamental cross sections $\sg$ (scattering, absorption and extinction) are determined as an infinite sum over polarizations ($p=E,M$) and all partial $\ell$-waves: 

\begin{equation}
\begin{split}
\sigma_{\rm sca} = \sum_{p,\ell} \sigma_{{\rm sca};p\ell} &  = \dfrac{\pi}{k_h^2} \sum_{p,\ell} \left(2\ell+1 \right) \left| \mathbb{T}_{p\ell} \right|^2 \ , \\ 
\sigma_{\rm abs} = \sum_{p,\ell} \sigma_{{\rm abs};p\ell} &  = \dfrac{\pi}{2k_h^2} \sum_{p,\ell} \left(2\ell+1 \right) \left( 1 - \left| 1 + 2 \mathbb{T}_{p\ell} \right|^2 \right)  \ , \\
\sigma_{\rm ext} = \sum_{p,\ell} \sigma_{{\rm ext};p\ell} & = - \dfrac{2\pi}{k_h^2} \sum_{p,\ell} \left(2\ell+1 \right) \Re\left(\mathbb{T}_{p\ell} \right) \ ,
\end{split}
\label{eq:crosssec}
\end{equation}
where 
$\Re$ takes the real part, and $\mathbb{T}_{p\ell}$ is found from Eq.~\eqref{dcrat}.

On using polarized scattering waves parallel and perpendicular to the scattering plane,

\begin{equation}
    \begin{split}
        S_\parallel (\vartheta) & = - \sum_\ell \dfrac{2\ell+1}{\ell(\ell+1)} \left[ \mathbb{T}_{E\ell} \frac{{\rm d} P_\ell^1 (\cos\vartheta)}{{\rm d}\vartheta} + \mathbb{T}_{M\ell} \frac{P_\ell^1 (\cos\vartheta)}{\sin\vartheta}  \right] \ , \\
        S_\perp (\vartheta) & = - \sum_\ell \dfrac{2\ell+1}{\ell(\ell+1)} \left[ \mathbb{T}_{M\ell} \frac{{\rm d} P_\ell^1 (\cos\vartheta)}{{\rm d}\vartheta} + \mathbb{T}_{E\ell} \frac{P_\ell^1 (\cos\vartheta)}{\sin\vartheta}  \right] \ ,
    \end{split}
    \label{eq:swaves}
\end{equation}
one can easily build up scattering matrix, find Stokes parameters~\cite[Eq. (4.77)]{Bohren1998}, and obtain angle-resolved scattering pattern.

\subsection{Near-field properties}
\label{sec:nearfld}
For a given illumination, and, thus, known expansion coefficients $A_{pL}(n)$ and $B_{pL}(n)$, the electromagnetic field can be unambiguously defined by Eqs.~\eqref{eq:Efld} and \eqref{eq:Hfld}.
Below we consider more sophisticated near-field properties.

\subsubsection{Electromagnetic energy}
\label{sec:energy}
Total electromagnetic energy $W$ stored within a multilayered sphere is a sum of energies, $W_n$, stored within any given $n$-th shell, which in turn is an integral of an electromagnetic energy density $w_n(r)$ over the $n$-th shell:

\begin{equation}
   W = \sum_{n=1}^{N} W_n = \sum_{n=1}^{N} \int_{r_{n-1}}^{r_n} w_n(r) r^2 {\rm d} r \ , \quad 
   w_n(r) = \dfrac{1}{4} \oint \left[ G_e(\veps_n) \left| \vE(\vr) \right|^2 + G_m(\mu_n)\left| \vH (\vr)\right|^2 \right] {\rm d} \Om \ ,
   \label{eq:wdef}
\end{equation}
where $r_0=0$, i.e. the core ``inner radius''.
Here we limit the discussion to a {\em plane wave} excitation.
For non-dispersive (usually dielectric) shells~\cite{Bott1987}, electric and magnetic coefficients in \eqref{eq:wdef} are
\begin{equation}
    G_e(\veps_n) = \Re(\veps_n) \ , \qquad G_m(\mu_n) = \Re(\mu_n) \ ,
    \label{eq:Gconst}
\end{equation}
while for the dispersive and absorbing shells~\cite{Loudon1970}:

\begin{equation}
    G_e (\veps_n,\om) =    \Re(\veps_n) + \dfrac{2\om}{\gamma_D} \Im(\veps_n) \ ,
    \label{eq:GLoudon}
\end{equation}
where $\gamma_D$ is the free electron damping constant in the Drude formula, and $\Im$ takes the imaginary part.
In our code, we use Drude fit parameters from either Blaber et al.~\cite{Blaber2009} or Ordal et al.~\cite{Ordal1985}.

Electric and magnetic components from Eq.~\eqref{eq:wdef} are nothing but the \textit{orientation-averaged} electric and magnetic field intensities:

\begin{equation}
\begin{split}
\oint \left| \vE \right|^2 \, {\rm d} \Om & = 2 \pi |E_0|^2 \sum_{\ell=1}^\infty \left[(2\ell+1)\, |\bar{f}_{M\ell}|^2 + (\ell+1)\, |\bar{f}_{E,\ell-1}|^2 + \ell\, |\bar{f}_{E,\ell+1}|^2\right] \ , 
\\
\oint \left| \vH \right|^2 \, {\rm d} \Om & = 2 \pi |E_0|^2  \frac{|\veps_n|}{|\mu_n|} \sum_{\ell=1}^\infty \left[ (2\ell+1)\, |\bar{f}_{E\ell}|^2 + (\ell+1)\, | \bar{f}_{M,\ell-1}|^2  + \ell\, |\bar{f}_{M,\ell+1}|^2 \right] \ .
\end{split}
\label{eq:eh2m}
\end{equation}
Here $\bar{f}_{p\ell,\ell\pm 1} = \bar{A}_{p\ell}(n) j_{\ell,\ell\pm 1}(k_n r) + \bar{B}_{p\ell}(n) h^{(1)}_{\ell,\ell\pm 1}(k_n r)$.
%
Note that in all cases the spherical Bessel functions of either $\ell$ and $\ell\pm 1$ orders are multiplied by the expansions coefficients $\bar{A}_{p\ell}$ and $\bar{B}_{p\ell}$ with the index $\ell$. 
These expansion coefficients are 
determined via raising ${\cal T}_{p\ell}(n)$ and lowering ${\cal M}_{p\ell}(n)$ composite transfer matrices as

\begin{equation}
\bar{A}_{p\ell}(n) =
\begin{cases} 
1 / {\cal T}_{11;p\ell} (N+1) \ , & n=1 \ , \\
{\cal M}_{11;p\ell} (n) + {\cal M}_{12;p\ell} (n) \dfrac{{\cal T}_{21;p\ell} (N+1)}{{\cal T}_{11;p\ell} (N+1)} \ , & 1<n<N+1 \ , \\
1 \ , & n = N+1 \ ,
\end{cases}
\end{equation}

\begin{equation}
\bar{B}_{p\ell}(n) =
\begin{cases}
0 \ , & n =1 \ , \\
{\cal M}_{21;p\ell} (n) + {\cal M}_{22;p\ell} (n) \dfrac{{\cal T}_{21;p\ell} (N+1)}{{\cal T}_{11;p\ell} (N+1)} \ , & 1<n<N+1 \ , \\
\dfrac{{\cal T}_{21;p\ell} (N+1)}{{\cal T}_{11;p\ell} (N+1)} \ , & n=N+1 \ .
\end{cases}    
\end{equation}
Because the surface integrals of electric and magnetic field intensities are performed analytically~\cite{Rasskazov19JOSAA}, the calculation of average intensity costs the same computational time as determining intensity at a {\em single} given point.

The radial integrations in \eqref{eq:wdef} are performed by using Lommel's integration formulas~\cite{Rasskazov19JOSAA}:

\begin{equation}
\begin{split}
\int_{r_{n-1}}^{r_n} r^2 {\rm d}r & \oint \left| \vE \right|^2 {\rm d}\Om = \\
& 2 \pi |E_0|^2 \frac{r^3}{x^2-x^{*2}}
\left. \sum_{\ell=1}^\infty \left[ 
(2\ell+1) \bar{F}_{M\ell} + (\ell+1) \bar{F}_{E,\ell-1} + \ell \bar{F}_{E,\ell+1}
\right]\right|_{r=r_{n-1}}^{r=r_n} \ , \\
\int_{r_{n-1}}^{r_n} r^2 {\rm d} r & \oint \left| \vH \right|^2 {\rm d}\Om = \\
& 2 \pi |E_0|^2 \frac{r^3}{x^2-x^{*2}} \frac{|\veps_n|}{|\mu_n|}
\left.\sum_{\ell=1}^\infty \left[ 
(2\ell+1) \bar{F}_{E\ell} + (\ell+1) \bar{F}_{M,\ell-1} + \ell \bar{F}_{M,\ell+1}
\right]\right|_{r=r_{n-1}}^{r=r_n} \ .
\end{split}
\label{eq:eh2mi}
\end{equation}
Here $x = k_n r$, and purely {\em imaginary} functions $\bar{F}_{p\ell} = 2i \Im\left(x \bar{f}_{p\ell+1}(x) \bar{f}^*_{p\ell}(x) \right)$ are cancelled by purely \textit{imaginary} $x^2-x^{*2}=4i \Re(x) \Im(x)$ in the denominator, which results in purely \textit{real} integrals in \eqref{eq:eh2mi}.
For a special case of \textit{lossless} shells, the denominator $x^2-x^{*2}$ vanishes.
It can be eliminated by using l'H\^{o}pital's rule~\cite{Rasskazov19JOSAA} to get the following amendments in Eq.~\eqref{eq:eh2mi}: $x^2-x^{*2} \to 2x$ and $\bar{F}_{p\ell}\ \to x \left(|\bar{f}_{p \ell}(x)|^2 + |\bar{f}_{p \ell+1}(x)|^2\right) 
         - (2\ell+1) \Re\left( \bar{f}_{p\ell}(x) \bar{f}^*_{p\ell+1}(x)\right)$.

Substitution of \eqref{eq:eh2m} and \eqref{eq:eh2mi} into \eqref{eq:wdef} yields in explicit expressions for the total electromagnetic energy $W_n$ stored within each shell of the multilayered sphere and for the electromagnetic energy density $w_n(r)$. 
These relations are general and valid for any shell, including the $(N+1)$-th shell being a surrounding medium.
For the comprehensive derivation of the equations above, we refer the Reader to the Ref.~\cite{Rasskazov19JOSAA}, which generalizes the pioneering work by Bott and Zdunkowski on homogeneous spheres~\cite{Bott1987} and subsequent follow-ups for magnetic spheres~\cite{Arruda2010} and core-shells~\cite{Arruda2012}.

\subsubsection{Spontaneous decay rates}
\label{sec:decay}
Radiative and nonradiative decay rates (normalized with respect to $\Gamma_{\rm rad;0}$, the intrinsic radiative decay rate in the absence of a multilayered sphere) for a dipole emitter located in $n_d$-th shell at $r_d$ distance from a center of a sphere (see Fig.~\ref{fig:scheme}) are given by~\cite{Moroz2005}:

\begin{equation}
\begin{split}
\dfrac{\Gm^{\perp}_{\rm rad}}{\Gm_{\rm rad;0}}
& = \dfrac{3}{2 x_d^4} {\cal N}_{\rm rad} \sum_{\ell=1}^{\infty}\ell(\ell+1)(2\ell+1) \left| {\cal F}_{E\ell}(x_d) \right|^2 \ , \\
\dfrac{\Gm^{\parallel}_{\rm rad}}{\Gm_{\rm rad;0}} 
& = \dfrac{3}{4 x_d^2} {\cal N}_{\rm rad} \sum_{\ell=1}^{\infty}(2\ell+1) \left[ \left| {\cal F}_{M\ell}(x_d) \right|^2 + \left| {\cal F}^{\;\prime}_{E\ell}(x_d) \right|^2 \right] \ , \\
\dfrac{\Gm^{\perp}_{\rm nrad}}{\Gm_{\rm rad;0}} 
& = \dfrac{3k^3_d}{2x_d^4} {\cal N}_{\rm nrad} \sum_{\Im(\veps_a)>0} \Im(\veps_a) \sum_{\ell=1}^{\infty} \ell(\ell+1)(2\ell+1) I_{E\ell;a} \left| {\cal D}_{E\ell;a}(x_d) \right|^2 \ , \\
\dfrac{\Gm^{\parallel}_{\rm nrad}}{\Gm_{\rm rad;0}} 
& = \dfrac{3k^3_d}{4x_d^2} {\cal N}_{\rm nrad} \sum_{\Im(\veps_a)>0} \Im(\veps_a) \sum_{\ell=1}^{\infty} (2\ell+1) \left[ I_{M\ell;a} \left| {\cal D}_{M\ell;a}(x_d) \right|^2 + I_{E\ell;a} \left| {\cal D}^{\prime}_{E\ell;a}(x_d) \right|^2 \right] \ ,
\end{split}
\label{eq:decay}
\end{equation}
where $x_d = k_d r_d $ and $k_d = 2\pi\eta_d/\ld$.
Coefficients ${\cal N}_{\rm rad}$ and ${\cal N}_{\rm nrad}$ depend on whether the decay rates were normalized with respect to the radiative decay rates in infinite homogeneous medium having the refractive index of (i) the \textit{host} or (ii) having the refractive index of the \textit{shell} where the dipole emitter is located:

\begin{equation}
    {\cal N}^{\rm host}_{\rm rad} = \dfrac{\eta_d^3}{\veps_d} \dfrac{\veps_h}{\eta^3_h} \ , \quad 
    {\cal N}^{\rm shell}_{\rm rad} = \left(\dfrac{\eta_d}{\eta_h}\right)^6 \left( \dfrac{\veps_h}{\veps_d} \right)^2 \ , \qquad
    {\cal N}^{\rm host}_{\rm nrad} = \dfrac{\eta_d^3}{\eta^3_h} \dfrac{\veps_h}{\veps_d^2} \ ,  \quad 
    {\cal N}^{\rm shell}_{\rm nrad} = \dfrac{1}{\veps_d} \cdot
    \label{eq:dcynorm}
\end{equation}
Functions ${\cal F}_{p\ell}(x_d)$ and ${\cal D}_{p\ell;a}(x_d)$ depend 
on the relative position of the emitter with respect to the sphere, and to a $n_a$-th absorbing shell:

\begin{equation}
    {\cal F}_{p\ell}(x_d) = 
    \begin{cases}
        \dfrac{\psi_\ell(x_d)}{{\cal M}_{21;p\ell}(1)} \ , & n_d=1 \ , \\[10pt]
        \dfrac{{\cal T}_{11;p\ell}(n_d)\psi_\ell(x_d) + {\cal T}_{21;p\ell}(n_d)\zeta_\ell(x_d)}{{\cal T}_{11;p\ell}(n_d){\cal M}_{22;p\ell}(n_d) - {\cal T}_{21;p\ell}(n_d){\cal M}_{12;p\ell}(n_d)} \ ,  
        & 1<n_d\leq N \ , \\[10pt]
        \psi_\ell (x_d)+\dfrac{{\cal T}_{21;p\ell}(N+1)}{{\cal T}_{11;p\ell}(N+1)} \zeta_\ell(x_d) \ , 
        & n_d=N+1 \ ,
    \end{cases}
\end{equation}

\begin{equation}
    {\cal D}_{p\ell;a}(x_d) = 
    \begin{cases}
        \dfrac{\psi_\ell(x_d)}{{\cal M}_{21;p\ell}(1)} \ , & n_d=1 \ , \\[10pt]
        \dfrac{{\cal T}_{11;p\ell}(n_d)\psi_\ell(x_d) + {\cal T}_{21;p\ell}(n_d)\zeta_\ell(x_d)}{{\cal T}_{11;p\ell}(n_d){\cal M}_{22;p\ell}(n_d) - {\cal T}_{21;p\ell}(n_d){\cal M}_{12;p\ell}(n_d)} \ , & 1<n_d<n_a \ , \\[10pt]
        \dfrac{{\cal M}_{22;p\ell}(n_d)\psi_\ell(x_d) + {\cal M}_{12;p\ell}(n_d)\zeta_\ell(x_d)}{{\cal T}_{11;p\ell}(n_d){\cal M}_{22;p\ell}(n_d) - {\cal T}_{21;p\ell}(n_d){\cal M}_{12;p\ell}(n_d)} \ , & n_a<n_d\leq N \ , \\[10pt]
        \zeta_\ell \left(x_d\right) \ , & n_d = N+1 \ .
    \end{cases}
\end{equation}
The summation over nonradiative decay channels in Eqs.~\eqref{eq:decay} includes each absorbing shell, i.e. shell with $\Im(\veps_a)>0$.
Respective volume integrals $I_{E\ell;a}$ and $I_{M\ell;a}$ are

\begin{equation}
    \begin{split}
        I_{M\ell;a} = & \dfrac{1}{|k_a|^2} \int_a \left| {\cal A}_{M\ell;a} \psi_\ell(k_a r) + {\cal B}_{M\ell;a} \zeta_\ell(k_a r)  \right|^2 {\rm d}r \ , \\
        I_{E\ell;a} = & \dfrac{\ell(\ell+1)}{|k_a|^4} \int_a \left| {\cal A}_{E\ell;a} \psi_\ell(k_a r) + {\cal B}_{E\ell;a} \zeta_\ell(k_a r)  \right|^2 \dfrac{{\rm d}r}{r^2} + \\
        & \dfrac{1}{|k_a|^2} \int_a \left| {\cal A}_{E\ell;a} \psi^{\prime}_{\ell}(k_a r) + {\cal B}_{E\ell;a} \zeta^{\prime}_{\ell}(k_a r) \right|^2 {\rm d}r \ .
    \end{split}
    \label{eq:Iabs}
\end{equation}
Here $k_a = 2\pi\eta_a/\ld$ is the wavevector in the absorbing shell, and coefficients ${\cal A}_{p\ell;a}$ and ${\cal B}_{p\ell;a}$ are

\begin{equation}
    {\cal A}_{p\ell;a} = 
    \begin{cases}
        1 \ , & n_a=1 \; \text{and} \; n_d<N+1 \ , \\[10pt]
        {\cal T}_{11;p\ell}(n_a) \ , & 1<n_a<n_d<N+1 \ , \\[10pt]
        {\cal M}_{12;p\ell}(n_a) \ , & n_d<n_a \ , \\[10pt]
        {\cal M}_{11;p\ell}(n_a) + {\cal M}_{12;p\ell}(n_a) \dfrac{{\cal T}_{21;p\ell}(N+1)}{{\cal T}_{11;p\ell}(N+1)} \ , & n_d=N+1 \ ,
    \end{cases}
\end{equation}

\begin{equation}
    {\cal B}_{p\ell;a} = 
    \begin{cases}
        0 \ , & n_a = 1 \ , \\[10pt]
        {\cal T}_{21;p\ell}(n_a) \ , & 1<n_a<n_d<N+1 \ , \\[10pt]
        {\cal M}_{22;p\ell}(n_a) \ , & n_d<n_a \ , \\[10pt]
        {\cal M}_{21;p\ell}(n_a) + {\cal M}_{22;p\ell}(n_a) \dfrac{{\cal T}_{21;p\ell}(N+1)}{{\cal T}_{11;p\ell}(N+1)}  \ , & n_a\neq 1 \; \text{and} \; n_d=N+1 \ .
    \end{cases}
\end{equation}
Note that ${\cal A}_{p\ell;a}$ and ${\cal B}_{p\ell;a}$ depend only on the \textit{relative} position of the dipole with respect to the absorbing shell.
We make use of this property to optimize calculations, namely, we get $I_{E\ell;a}$ and $I_{M\ell;a}$ once for a set of $r_d$, considering only the \textit{relative} position of the dipole emitter.

For the detailed discussion of the spontaneous decay rates of a dipole emitter in a presence of a general multilayered sphere, we refer the reader to Ref.~\cite{Moroz2005}.

\subsection{Convergence criteria}
\label{sec:conv}
Numerical implementation of equations above, which involve infinite summation over $\ell$, requires truncation at some finite number $\ell_{\rm max}$.
For far-field properties, Eqs.~\eqref{eq:crosssec} and \eqref{eq:swaves}, the classic choice is the Wiscombe criterion~\cite{Wiscombe1980}:

\begin{equation}
\ell_{\rm max} =
\begin{cases}
x_N + 4x_N^{1/3} + 1 \ , \qquad 0.02 \leq x_N \leq 8 \ , \\
x_N + 4.05x_N^{1/3} + 2 \ , \qquad 8 < x_N \ ,
\end{cases}    
\label{eq:wsbcrt}
\end{equation}
where $x_N=k_h r_N $ is the usual size parameter. 
Keeping in mind plasmonics and nanophotonics applications as intended for our package, and recalling finite precision of the MATLAB, we limit the discussion for multilayered spheres with $x_N<50$ requiring $\ell_{\rm max}\lesssim 70$, which is more than sufficient.

For the near-field, Eqs.~\eqref{eq:Efld}, \eqref{eq:Hfld}, and for the electromagnetic energy, Eqs.~\eqref{eq:wdef}, \eqref{eq:eh2m}, \eqref{eq:eh2mi}, slightly different cut-off is recommended~\cite{Allardice2014}:

\begin{equation}
\ell_{\rm max} = x_N + 11x_N^{1/3} + 1 \ . 
\end{equation}

For the dipole source implied in Eq.~\eqref{eq:decay}, especially when considering non-radiative decay rates~\cite{Moroz2005,Moroz2010}, the \textit{accuracy} has to be set instead of $\ell_{\rm max}$, since there are no general guidelines for choosing the latter in this case.

\subsection{Electron free path correction}
\label{sec:freepath}
For thin metallic shell with thickness $(r_n - r_{n-1})$ less than a free electron path, the bulk permittivity $\veps_{n,{\rm bulk}}$ has to be corrected to take into account electron scattering from the shell surface~\cite{Moroz2008}: 

\begin{equation}
    \veps_n \to \varepsilon_{n,{\rm bulk}} + \dfrac{\omega^2_p}{\omega^2 + i \gamma_D \omega} - \dfrac{\omega^2_p}{\omega^2 + i \gamma \omega} \ , \quad     \gamma = \gamma_D + \dfrac{\upsilon_F}{L_{\rm eff}} \ , \quad  L_{\rm eff} = \dfrac{4}{3}\dfrac{r_n^3 - r_{n-1}^{3}}{r_n^2 + r_{n-1}^{2}} \ ,
    \label{eq:elfrpth}
\end{equation}
where $\upsilon_F$ is the Fermi velocity and $\om_p$ is plasma frequency.
This modification of thin metallic shells permittivity may have crucial consequences for their electromagnetic properties~\cite{Averitt1999,Ruppin2015,Zakomirnyi17JQSRT,Kostyukov19JQSRT,Zakomirnyi20PCCP}.

\section{Computer code}
\label{sec:code}

\subsection{Overview}
Fundamental properties are calculated with the following functions:
\begin{itemize}
    \item \verb|t_mat.m| calculates transfer matrices with Eqs.~\eqref{eq:bacmtm}~--~\eqref{eq:foretm} and returns their ordered products \eqref{eq:ordprod}.
    Any other function which returns electromagnetic properties (decay rates, electromagnetic energy and fields, far-field properties and etc) makes use of pre-calculated ordered products of transfer matrices for a better performance;
    \item \verb|decay.m| returns decay rates calculated with Eqs.~\eqref{eq:decay}.
    Decay rates are normalized (see Eqs.~\eqref{eq:dcynorm}) with respect to radiative decay rates of a dipole in a homogeneous medium (without a multilayered sphere) with the refractive index of a \textit{shell} where the dipole is embedded, $\eta_d$, or with the refractive index of a \textit{host}, $\eta_h$;
    \item \verb|nrg_dens.m| returns normalized by the factor $\pi |E_0|^2 \veps_h$ \cite[Sec.3E]{Rasskazov19JOSAA} electric and magnetic components of electromagnetic energy density, $w_n(r)$, given by Eq.~\eqref{eq:wdef}, and orientation-averaged intensities of electric and magnetic fields given by Eq.~\eqref{eq:eh2m};
    \item \verb|nrg_tot.m| returns normalized by the factor $(2/3)\pi |E_0|^2 (r_n - r_{n-1}^3) \veps_h$ \cite[Sec.3E]{Rasskazov19JOSAA} electric and magnetic components of \textit{total} electromagnetic energy $W_n$ stored within $n$-th shell given by Eq.~\eqref{eq:wdef};
    \item \verb|near_fld.m| returns electric, $\vE(\vr) = \vE_E(\vr) + \vE_M(\vr)$, and magnetic, $\vH(\vr) = \vH_E(\vr) + \vH_M(\vr)$ near-field distributions given by Eqs.~\eqref{eq:Efld} and \eqref{eq:Hfld}.
    We take the advantage of the spherical symmetry of the problem and pre-calculate computationally expensive $\vr$-dependent Bessel functions and $\cos\theta$-dependent associated Legendre polynomials only for \textit{unique} values of $\vr$ and $\cos\theta$.
    Usually, the amount of these \textit{unique} $\vr$ and $\cos\theta$ is significantly smaller than the respective \textit{total} number of points in the rectangular mesh, which results in faster execution;
    \item \verb|far_fld.m| returns polarized scattering waves from Eq.~\eqref{eq:swaves};
    \item \verb|crs_sec.m| returns fundamental cross sections from Eq.~\eqref{eq:crosssec};
    \item \verb|el_fr_pth.m| returns corrected permittivity according to Eq.~\eqref{eq:elfrpth}.
\end{itemize}
Suitable combinations or minor post-processing of these fundamental characteristics may be used for almost every known application of multilayered spheres briefly discussed in Sec.~\ref{sec:disc} below.

\subsection{Verification and performance}
\begin{figure}
    \centering
    \includegraphics{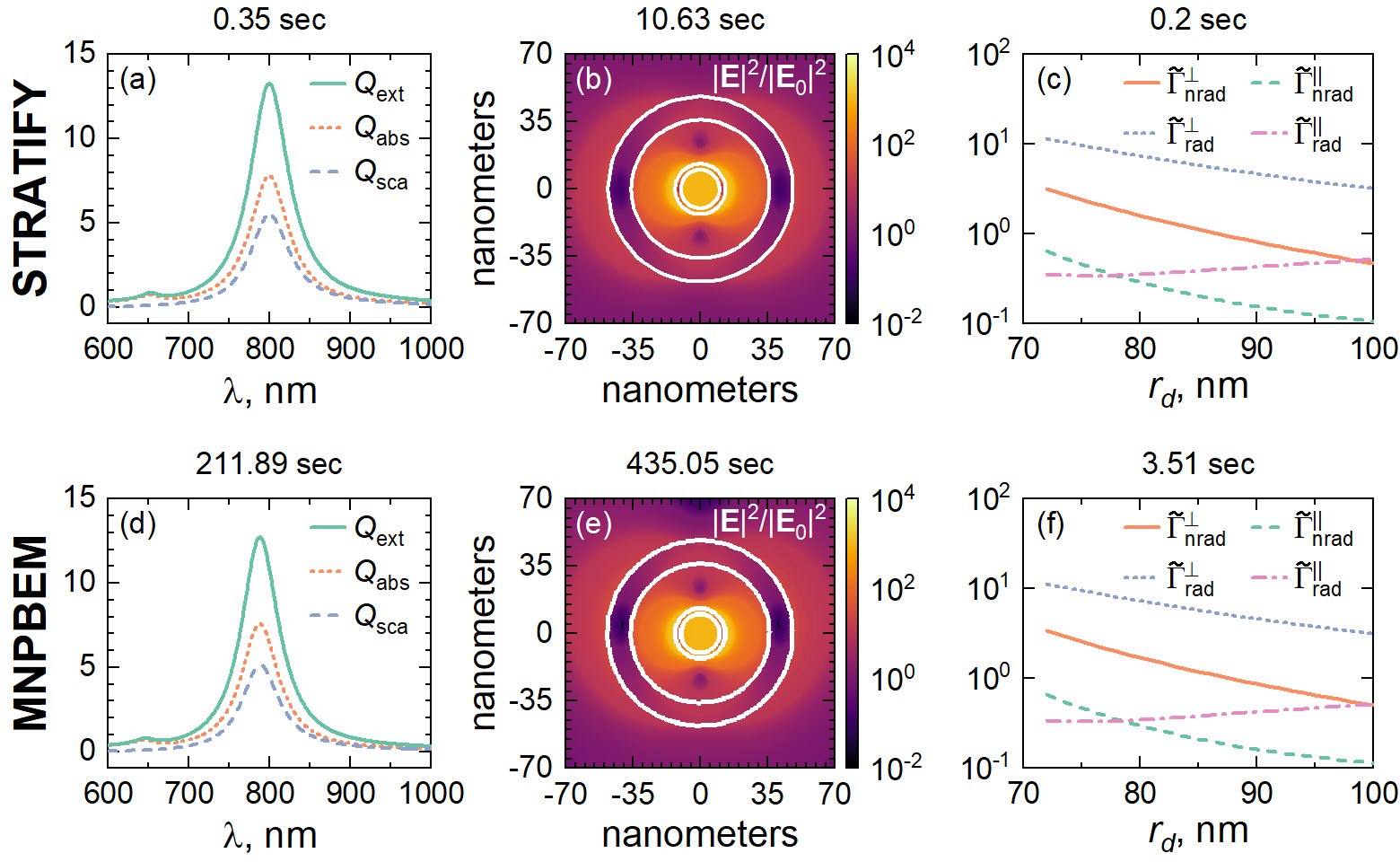}
    \caption{Comparison between our code (top, STRATIFY) and freely available BEM-based package (bottom, MNPBEM). 
    (a),(d)~Normalized fundamental cross-section ($Q=\sigma/\pi r_2^2$) of SiO${}_2$@Au core-shell sphere with $\{r_1,r_2\}=\{50,55\}$nm, in air host; 
    (b),(e) intensity of the electric field for SiO${}_2$@Au@SiO${}_2$@Au matryoshka with $\{r_1,r_2,r_3,r_4\}=\{10,13,36,48\}$~nm, in water host, at $\ld=690$~nm~\cite[cf. Fig.3b]{Meng2017};
    (c),(f) normalized spontaneous decay rates ($\tilde{\Gm} = \Gm / \Gm_{\rm rad;0}$) for the dipole emitter located at $r_d$ distance from the center of Au@SiO${}_2$ core-shell sphere with $\{r_1,r_2\}=\{50,70\}$~nm, in water host, at $\ld=614$~nm.
    Run times of codes on a laptop with 2.6 GHz 6-Core Intel Core i7 processor are shown on top of each plot.
    For a fair comparison, the same number of points for $\ld$ (801) in (a),(c), same $700\times 700$ mesh in (b),(e), and same number of points for $r_d$ (29) in (c),(f) are used. 
    Note typically more than $50$ times faster speed of our package compared to the freely available BEM-based package.}
    \label{fig:bnchmrk}
\end{figure}

We have compared STRATIFY with freely available exact BEM-based~\cite{GarciadeAbajo2002} solver MNPBEM~\cite{Hohenester2012,Hohenester2014,Waxenegger2015,Hohenester2018}.
The latter package has been chosen for a benchmark since it is a freely available comprehensive MATLAB code capable of calculating most of the quantities considered in our code.
For testing purposes, we have considered fundamental cross-sections, electric near-field distribution, and spontaneous decay rates of dipole emitter in the presence of matryoshkas composed of different combinations of Au and SiO${}_2$ layers.
It can be easily seen from Fig.~\ref{fig:bnchmrk} that our code produces the same results as the exact BEM method, but for sufficiently lower computational price.

\section{Discussion and Conclusions}
\label{sec:disc}
The developed package is ready to be applied in a number of well-established applications of multilayered spheres in optics and photonics.
Below we discuss the most common examples.
Due to enormous electric field enhancement in multilayered metal-dielectric nanospheres, they are considered as good candidates for a number of applications.
Squared electric field intensities from Eq.~\eqref{eq:eh2m} or Eq.~\eqref{eq:eh2mi} are measure for performance of nanostructures in SERS~\cite{Chew1976,Kodali2010,Kodali2010a,Lim2011,Khlebtsov2017,Li2017a} and second harmonic generation~\cite{Pu2010,Scherbak2018}.
For fluorescence or upconversion enhancement, where the spontaneous decay rates of the dipole emitters are modified, the generic enhancement factor is nothing but a product of excitation rate enhancement (at excitation wavelength, $\ld_{\rm exc}$) and quantum yield (at emission wavelength, $\ld_{\rm ems}$): 

\begin{equation}
    F = \dfrac{\gamma_{\rm exc}}{\gamma_{\rm exc;0}} \dfrac{q_{\rm ems}}{q_{\rm ems;0}} \ , \qquad 
        q_{\rm ems} = \dfrac{\Gm_{\rm rad}/\Gm_{{\rm rad};0}}{\Gm_{\rm rad}/\Gm_{{\rm rad};0} + \Gm_{\rm nrad}/\Gm_{{\rm rad};0} + (1-q_{\rm ems;0})/q_{\rm ems;0}} \cdot
    \label{eq:genenh}
\end{equation}
Here $\gm_{\rm exc}/\gm_{\rm exc;0} \propto |{\bf E}|^{2\mathcal{P}}/|{\bf E}_0|^{2\mathcal{P}}$, $\mathcal{P}$ is the number of photons involved in the excitation process ($\mathcal{P}=1$ for fluorescence enhancement, $\mathcal{P}=2$ for the upconversion), $q_{\rm ems;0}$ is the intrinsic quantum yield of the emitter, and 
$\Gm_{\rm rad, nrad} = (2\Gm^\parallel_{\rm rad, nrad} + \Gm^\perp_{\rm rad, nrad})/3$ is
an orientationally averaged decay rate determined at a fixed dipole radial position by averaging over all possible orientations of a dipole emitter. 
Orientationnaly averaged fields in $\gm_{\rm exc}$, and orientationally averaged decay rates, $\Gm_{\rm rad, nrad}$, are reasonable choice for reliable estimates, 
unless, of course, the emitter is positioned with a controllable orientation of its electric dipole moment at a particular (e.g. a hot spot) location.

Cloaking, Kerker effect~\cite{Lee2018},
super- or optimally tuned scattering~\cite{Ruan2010,Ruan2011,Argyropoulos2013,Fleury2014,Sheverdin2019,Lepeshov2019,Yezekyan2020} and absorption~\cite{Tuersun2013,Fleury2014,Ladutenko2015,Xue2016,Zakomirnyi17JQSRT,Sheverdin2019,Yezekyan2020},
embedded photonic eigenvalues~\cite{Monticone2014}, spasing~\cite{Gordon2007,Calander2012,Passarelli2016,Pezzi2019} and other intriguing phenomena~\cite{Miroshnichenko2010a} are easily understood from fundamental cross sections~\eqref{eq:crosssec} and scattering patterns~\eqref{eq:swaves}.

We have summarized a self-consistent and comprehensive RTMM theory reported earlier 
in our \cite{Moroz2005,Rasskazov19JOSAA} for electromagnetic light scattering from general multilayered spheres composed of isotropic shells.
Within the framework of RTMM, we have developed an efficient multi-purpose MATLAB package for calculating fundamental properties of multilayered spheres.
Our package is one-of-a-kind freely available software which allows for a simultaneous calculation of a wide range of electromagnetic properties and is ready to be used for a broad number of applications in chemistry, optics and photonics, including optimization problems and machine-learning studies. 
We hope that the generalization presented here and corresponding MATLAB code will serve as a useful tool for photonics, physics, chemistry and other scientific communities and will boost the researches involving various kinds of multilayered spheres.

Extensions of our code for ultra-thin metallic shell characterized by nonlocal dielectric functions~\cite{Melnyk1970,Anderegg1971,Ruppin1975,Leung1990,Rojas1988,David2011,Huang2014,Mortensen2014,Dong2017c,Eremin2019}, an optically active shells~\cite{Bohren1974,Bohren1975,Lakhtakia1985,Engheta1990,Yokota2001,Guzatov2012a,Klimov2012,Arruda2013a}, or including perfectly conducting boundary conditions at the sphere core are, following the theory developed in Ref.~\cite{Moroz2005}, rather straightforward. 
Further generalization of the package may include illumination with focused, Gaussian, or other beams~\cite{Gouesbet1988,Onofri1995,Li2007,Mojarad2009}. 
An incorporation of \textit{magnetic} dipole emitters~\cite{DeDood2001,DeDood2001a,Wiecha2018} or modeling the effect of a multilayered sphere on the far-field radiation directivity of a dipole antenna should follow soon.



\begin{thebibliography}{100}
\newcommand{\enquote}[1]{``#1''}

\bibitem{Neeves1988}
A.~E. Neeves and M.~H. Birnboim, \enquote{{Composite structures for the
  enhancement of nonlinear optical materials},} {\protect\JournalTitle{Optics
  Letters}} \textbf{13}, 1087--1089 (1988).

\bibitem{Neeves1989}
A.~E. Neeves and M.~H. Birnboim, \enquote{{Composite structures for the
  enhancement of nonlinear-optical susceptibility},}
  {\protect\JournalTitle{Journal of the Optical Society of America B}}
  \textbf{6}, 787--796 (1989).

\bibitem{Zhou1994}
H.~S. Zhou, I.~Honma, H.~Komiyama, and J.~W. Haus, \enquote{{Controlled
  synthesis and quantum-size effect in gold-coated nanoparticles},}
  {\protect\JournalTitle{Physical Review B}} \textbf{50}, 12052--12056 (1994).

\bibitem{Averitt1997}
R.~D. Averitt, D.~Sarkar, and N.~J. Halas, \enquote{{Plasmon resonance shifts
  of Au-coated Au2S nanoshells: Insight into multicomponent nanoparticle
  growth},} {\protect\JournalTitle{Physical Review Letters}} \textbf{78},
  4217--4220 (1997).

\bibitem{Oldenburg1998}
S.~Oldenburg, R.~Averitt, S.~Westcott, and N.~Halas, \enquote{{Nanoengineering
  of optical resonances},} {\protect\JournalTitle{Chemical Physics Letters}}
  \textbf{288}, 243--247 (1998).

\bibitem{Graf2002}
C.~Graf and A.~van Blaaderen, \enquote{{Metallodielectric colloidal core-shell
  particles for photonic applications},} {\protect\JournalTitle{Langmuir}}
  \textbf{18}, 524--534 (2002).

\bibitem{Hasegawa2006}
K.~Hasegawa, C.~Rohde, and M.~Deutsch, \enquote{{Enhanced surface-plasmon
  resonance absorption in metal-dielectric-metal layered microspheres},}
  {\protect\JournalTitle{Optics Letters}} \textbf{31}, 1136--1138 (2006).

\bibitem{Pu2010}
Y.~Pu, R.~Grange, C.-L. Hsieh, and D.~Psaltis, \enquote{{Nonlinear optical
  properties of core-shell nanocavities for enhanced second-harmonic
  generation},} {\protect\JournalTitle{Physical Review Letters}} \textbf{104},
  207402 (2010).

\bibitem{Scherbak2018}
S.~A. Scherbak and A.~A. Lipovskii, \enquote{{Understanding the second-harmonic
  generation enhancement and behavior in metal core-dielectric shell
  nanoparticles},} {\protect\JournalTitle{Journal of Physical Chemistry C}}
  \textbf{122}, 15635--15645 (2018).

\bibitem{Moroz1999}
A.~Moroz and C.~Sommers, \enquote{{Photonic band gaps of three-dimensional
  face-centred cubic lattices},} {\protect\JournalTitle{J. Phys.: Condens.
  Mat.}} \textbf{11}, 997--1008 (1999).

\bibitem{Zhang2000}
W.~Y. Zhang, X.~Y. Lei, Z.~L. Wang, D.~G. Zheng, W.~Y. Tam, C.~T. Chan, and
  P.~Sheng, \enquote{{Robust photonic band gap from tunable scatterers},}
  {\protect\JournalTitle{Physical Review Letters}} \textbf{84}, 2853--2856
  (2000).

\bibitem{Moroz2000}
A.~Moroz, \enquote{{Photonic crystals of coated metallic spheres},}
  {\protect\JournalTitle{Europhysics Letters (EPL)}} \textbf{50}, 466--472
  (2000).

\bibitem{Velikov2002}
K.~P. Velikov, A.~Moroz, and A.~Van~Blaaderen, \enquote{{Photonic crystals of
  core-shell colloidal particles},} {\protect\JournalTitle{Applied Physics
  Letters}} \textbf{80}, 49--51 (2002).

\bibitem{Moroz2002}
A.~Moroz, \enquote{{Metallo-dielectric diamond and zinc-blende photonic
  crystals},} {\protect\JournalTitle{Physical Review B}} \textbf{66}, 115109
  (2002).

\bibitem{Raschke2004}
G.~Raschke, S.~Brogl, A.~S. Susha, A.~L. Rogach, T.~A. Klar, J.~Feldmann,
  B.~Fieres, N.~Petkov, T.~Bein, A.~Nichtl, and K.~K{\"{u}}rzinger,
  \enquote{{Gold nanoshells improve single nanoparticle molecular sensors},}
  {\protect\JournalTitle{Nano Letters}} \textbf{4}, 1853--1857 (2004).

\bibitem{Jain2007}
P.~K. Jain and M.~A. El-Sayed, \enquote{{Surface plasmon resonance sensitivity
  of metal nanostructures: Physical basis and universal scaling in metal
  nanoshells},} {\protect\JournalTitle{Journal of Physical Chemistry C}}
  \textbf{111}, 17451--17454 (2007).

\bibitem{Ochsenkuhn2009}
M.~A. Ochsenk{\"{u}}hn, P.~R.~T. Jess, H.~Stoquert, K.~Dholakia, and C.~J.
  Campbell, \enquote{{Nanoshells for surface-enhanced Raman spectroscopy in
  Eukaryotic cells: Cellular response and sensor development},}
  {\protect\JournalTitle{ACS Nano}} \textbf{3}, 3613--3621 (2009).

\bibitem{Hirsch2003}
L.~R. Hirsch, R.~J. Stafford, J.~A. Bankson, S.~R. Sershen, B.~Rivera, R.~E.
  Price, J.~D. Hazle, N.~J. Halas, and J.~L. West, \enquote{{Nanoshell-mediated
  near-infrared thermal therapy of tumors under magnetic resonance guidance},}
  {\protect\JournalTitle{Proceedings of the National Academy of Sciences}}
  \textbf{100}, 13549--13554 (2003).

\bibitem{Ayala-Orozco2014}
C.~Ayala-Orozco, C.~Urban, M.~W. Knight, A.~S. Urban, O.~Neumann, S.~W.
  Bishnoi, S.~Mukherjee, A.~M. Goodman, H.~Charron, T.~Mitchell, M.~Shea,
  R.~Roy, S.~Nanda, R.~Schiff, N.~J. Halas, and A.~Joshi, \enquote{{Au
  nanomatryoshkas as efficient near-infrared photothermal transducers for
  cancer treatment: benchmarking against nanoshells},}
  {\protect\JournalTitle{ACS Nano}} \textbf{8}, 6372--6381 (2014).

\bibitem{Zakomirnyi17JQSRT}
V.~I. Zakomirnyi, I.~L. Rasskazov, S.~V. Karpov, and S.~P. Polyutov,
  \enquote{{New ideally absorbing Au plasmonic nanostructures for biomedical
  applications},} {\protect\JournalTitle{Journal of Quantitative Spectroscopy
  and Radiative Transfer}} \textbf{187}, 54--61 (2017).

\bibitem{Kostyukov19JQSRT}
A.~S. Kostyukov, A.~E. Ershov, V.~S. Gerasimov, S.~A. Filimonov, I.~L.
  Rasskazov, and S.~V. Karpov, \enquote{{Super-efficient laser hyperthermia of
  malignant cells with core-shell nanoparticles based on alternative plasmonic
  materials},} {\protect\JournalTitle{Journal of Quantitative Spectroscopy and
  Radiative Transfer}} \textbf{236}, 106599 (2019).

\bibitem{Phan2018a}
A.~D. Phan, N.~B. Le, N.~T.~H. Lien, and K.~Wakabayashi, \enquote{{Multilayered
  plasmonic nanostructures for solar energy harvesting},}
  {\protect\JournalTitle{Journal of Physical Chemistry C}} \textbf{122},
  19801--19806 (2018).

\bibitem{Wang2018}
Z.~Wang, X.~Quan, Z.~Zhang, and P.~Cheng, \enquote{{Optical absorption of
  carbon-gold core-shell nanoparticles},} {\protect\JournalTitle{Journal of
  Quantitative Spectroscopy and Radiative Transfer}} \textbf{205}, 291--298
  (2018).

\bibitem{Xu2020}
X.~Xu, A.~Dutta, J.~Khurgin, A.~Wei, V.~M. Shalaev, and A.~Boltasseva,
  \enquote{{TiN@TiO2 core-shell nanoparticles as plasmon-enhanced
  photosensitizers: The role of hot electron injection},}
  {\protect\JournalTitle{Laser {\&} Photonics Reviews}} \textbf{14}, 1900376
  (2020).

\bibitem{Chew1976}
H.~Chew, P.~J. McNulty, and M.~Kerker, \enquote{{Model for Raman and
  fluorescent scattering by molecules embedded in small particles},}
  {\protect\JournalTitle{Physical Review A}} \textbf{13}, 396--404 (1976).

\bibitem{Tovmachenko2006}
O.~G. Tovmachenko, C.~Graf, D.~J. van~den Heuvel, A.~van Blaaderen, and H.~C.
  Gerritsen, \enquote{{Fluorescence enhancement by metal-core/silica-shell
  nanoparticles},} {\protect\JournalTitle{Advanced Materials}} \textbf{18},
  91--95 (2006).

\bibitem{Zhang2006}
J.~Zhang, I.~Gryczynski, Z.~Gryczynski, and J.~R. Lakowicz,
  \enquote{{Dye-labeled silver nanoshell-bright particle},}
  {\protect\JournalTitle{Journal of Physical Chemistry B}} \textbf{110},
  8986--8991 (2006).

\bibitem{Ayala-Orozco2014a}
C.~Ayala-Orozco, J.~G. Liu, M.~W. Knight, Y.~Wang, J.~K. Day, P.~Nordlander,
  and N.~J. Halas, \enquote{{Fluorescence enhancement of molecules inside a
  gold nanomatryoshka},} {\protect\JournalTitle{Nano Letters}} \textbf{14},
  2926--2933 (2014).

\bibitem{Sakamoto2017}
N.~Sakamoto, T.~Onodera, T.~Dezawa, Y.~Shibata, and H.~Oikawa, \enquote{{Highly
  enhanced emission of visible light from core-dual-shell-type hybridized
  nanoparticles},} {\protect\JournalTitle{Particle {\&} Particle Systems
  Characterization}} \textbf{34}, 1700258 (2017).

\bibitem{Sun2020}
S.~Sun, I.~L. Rasskazov, P.~S. Carney, T.~Zhang, and A.~Moroz,
  \enquote{{Critical role of shell in enhanced fluorescence of metal-dielectric
  core-shell nanoparticles},} {\protect\JournalTitle{Journal of Physical
  Chemistry C}} p. acs.jpcc.0c03415 (2020).

\bibitem{Naiki2017}
H.~Naiki, H.~Oikawa, and S.~Masuo, \enquote{{Modification of emission photon
  statistics from single quantum dots using metal/SiO2 core/shell
  nanostructures},} {\protect\JournalTitle{Photochemical {\&} Photobiological
  Sciences}} \textbf{16}, 489--498 (2017).

\bibitem{Zhang2010a}
F.~Zhang, G.~B. Braun, Y.~Shi, Y.~Zhang, X.~Sun, N.~O. Reich, D.~Zhao, and
  G.~Stucky, \enquote{{Fabrication of Ag@SiO2@Y2O3:Er nanostructures for
  bioimaging: Tuning of the upconversion fluorescence with silver
  nanoparticles},} {\protect\JournalTitle{Journal of the American Chemical
  Society}} \textbf{132}, 2850--2851 (2010).

\bibitem{Priyam2012}
A.~Priyam, N.~M. Idris, and Y.~Zhang, \enquote{{Gold nanoshell coated
  NaYF4nanoparticles for simultaneously enhanced upconversion fluorescence and
  darkfield imaging},} {\protect\JournalTitle{J. Mater. Chem.}} \textbf{22},
  960--965 (2012).

\bibitem{Yuan2012a}
P.~Yuan, Y.~H. Lee, M.~K. Gnanasammandhan, Z.~Guan, Y.~Zhang, and Q.-H. Xu,
  \enquote{{Plasmon enhanced upconversion luminescence of NaYF4:Yb,Er@SiO2@Ag
  core-shell nanocomposites for cell imaging},}
  {\protect\JournalTitle{Nanoscale}} \textbf{4}, 5132--5137 (2012).

\bibitem{Kannan2013}
P.~Kannan, F.~A. Rahim, X.~Teng, R.~Chen, H.~Sun, L.~Huang, and D.-H. Kim,
  \enquote{{Enhanced emission of NaYF4:Yb,Er/Tm nanoparticles by selective
  growth of Au and Ag nanoshells},} {\protect\JournalTitle{RSC Advances}}
  \textbf{3}, 7718 (2013).

\bibitem{Ding2014}
Y.~Ding, X.~Zhang, H.~Gao, S.~Xu, C.~Wei, and Y.~Zhao, \enquote{{Plasmonic
  enhanced upconversion luminescence of {$\beta$}-NaYF4:Yb3+/Er3+ with Ag@SiO2
  core-shell nanoparticles},} {\protect\JournalTitle{Journal of Luminescence}}
  \textbf{147}, 72--76 (2014).

\bibitem{Xu2014}
W.~Xu, X.~Min, X.~Chen, Y.~Zhu, P.~Zhou, S.~Cui, S.~Xu, L.~Tao, and H.~Song,
  \enquote{{Ag-SiO2-Er2O3 nanocomposites: Highly effective upconversion
  luminescence at high power excitation and high temperature},}
  {\protect\JournalTitle{Scientific Reports}} \textbf{4}, 5087 (2014).

\bibitem{Qin2016}
Y.~Qin, Z.~Dong, D.~Zhou, Y.~Yang, X.~Xu, and J.~Qiu, \enquote{{Modification on
  populating paths of {$\beta$}-NaYF{\_}4:Nd/Yb/Ho@SiO{\_}2@Ag
  core/double-shell nanocomposites with plasmon enhanced upconversion
  emission},} {\protect\JournalTitle{Optical Materials Express}} \textbf{6},
  1942 (2016).

\bibitem{Wang2016a}
Z.~Wang, W.~Gao, R.~Wang, J.~Shao, Q.~Han, C.~Wang, J.~Zhang, T.~Zhang,
  J.~Dong, and H.~Zheng, \enquote{{Influence of SiO2 layer on the plasmon
  quenched upconversion luminescence emission of core-shell NaYF4:Yb,Er@SiO2@Ag
  nanocomposites},} {\protect\JournalTitle{Materials Research Bulletin}}
  \textbf{83}, 515--521 (2016).

\bibitem{Rasskazov18OMEx}
I.~L. Rasskazov, L.~Wang, C.~J. Murphy, R.~Bhargava, and P.~S. Carney,
  \enquote{{Plasmon-enhanced upconversion: engineering enhancement and
  quenching at nano and macro scales},} {\protect\JournalTitle{Optical
  Materials Express}} \textbf{8}, 3787--3804 (2018).

\bibitem{Lim2011}
D.-K. Lim, K.-S. Jeon, J.-H. Hwang, H.~Kim, S.~Kwon, Y.~D. Suh, and J.-M. Nam,
  \enquote{{Highly uniform and reproducible surface-enhanced Raman scattering
  from DNA-tailorable nanoparticles with 1-nm interior gap},}
  {\protect\JournalTitle{Nature Nanotechnology}} \textbf{6}, 452--460 (2011).

\bibitem{Li2017a}
J.-F. Li, Y.-J. Zhang, S.-Y. Ding, R.~Panneerselvam, and Z.-Q. Tian,
  \enquote{{Core-shell nanoparticle-enhanced Raman spectroscopy},}
  {\protect\JournalTitle{Chemical Reviews}} \textbf{117}, 5002--5069 (2017).

\bibitem{Noginov2009}
M.~A. Noginov, G.~Zhu, A.~M. Belgrave, R.~Bakker, V.~M. Shalaev, E.~E.
  Narimanov, S.~Stout, E.~Herz, T.~Suteewong, and U.~Wiesner,
  \enquote{{Demonstration of a spaser-based nanolaser},}
  {\protect\JournalTitle{Nature}} \textbf{460}, 1110--1112 (2009).

\bibitem{Calander2012}
N.~Calander, D.~Jin, and E.~M. Goldys, \enquote{{Taking plasmonic core-shell
  nanoparticles toward laser threshold},} {\protect\JournalTitle{Journal of
  Physical Chemistry C}} \textbf{116}, 7546--7551 (2012).

\bibitem{Baranov2013}
D.~G. Baranov, E.~Andrianov, A.~P. Vinogradov, and A.~A. Lisyansky,
  \enquote{{Exactly solvable toy model for surface plasmon amplification by
  stimulated emission of radiation},} {\protect\JournalTitle{Optics Express}}
  \textbf{21}, 10779--10791 (2013).

\bibitem{Arnold2016}
N.~Arnold, C.~Hrelescu, and T.~A. Klar, \enquote{{Minimal spaser threshold
  within electrodynamic framework: Shape, size and modes},}
  {\protect\JournalTitle{Annalen der Physik}} \textbf{528}, 295--306 (2016).

\bibitem{Passarelli2016}
N.~Passarelli, R.~A. Bustos-Mar{\'{u}}n, and E.~A. Coronado, \enquote{{Spaser
  and optical amplification conditions in gold-coated active nanoparticles},}
  {\protect\JournalTitle{Journal of Physical Chemistry C}} \textbf{120},
  24941--24949 (2016).

\bibitem{Galanzha2017}
E.~I. Galanzha, R.~Weingold, D.~A. Nedosekin, M.~Sarimollaoglu, J.~Nolan,
  W.~Harrington, A.~S. Kuchyanov, R.~G. Parkhomenko, F.~Watanabe, Z.~Nima,
  A.~S. Biris, A.~I. Plekhanov, M.~I. Stockman, and V.~P. Zharov,
  \enquote{{Spaser as a biological probe},} {\protect\JournalTitle{Nature
  Communications}} \textbf{8}, 15528 (2017).

\bibitem{Alu2008}
A.~Al{\`{u}} and N.~Engheta, \enquote{{Multifrequency optical invisibility
  cloak with layered plasmonic shells},} {\protect\JournalTitle{Physical Review
  Letters}} \textbf{100}, 113901 (2008).

\bibitem{Monticone2013}
F.~Monticone, C.~Argyropoulos, and A.~Al{\`{u}}, \enquote{{Multilayered
  plasmonic covers for comblike scattering response and optical tagging},}
  {\protect\JournalTitle{Physical Review Letters}} \textbf{110}, 113901 (2013).

\bibitem{Sheverdin2019}
A.~Sheverdin and C.~Valagiannopoulos, \enquote{{Core-shell nanospheres under
  visible light: Optimal absorption, scattering, and cloaking},}
  {\protect\JournalTitle{Physical Review B}} \textbf{99}, 075305 (2019).

\bibitem{Tsakmakidis2019}
K.~L. Tsakmakidis, O.~Reshef, E.~Almpanis, G.~P. Zouros, E.~Mohammadi,
  D.~Saadat, F.~Sohrabi, N.~Fahimi-Kashani, D.~Etezadi, R.~W. Boyd, and
  H.~Altug, \enquote{{Ultrabroadband 3D invisibility with fast-light cloaks},}
  {\protect\JournalTitle{Nature Communications}} \textbf{10}, 4859 (2019).

\bibitem{Hirsch2006}
L.~R. Hirsch, A.~M. Gobin, A.~R. Lowery, F.~Tam, R.~A. Drezek, N.~J. Halas, and
  J.~L. West, \enquote{{Metal nanoshells},} {\protect\JournalTitle{Annals of
  Biomedical Engineering}} \textbf{34}, 15--22 (2006).

\bibitem{Jankiewicz2012}
B.~Jankiewicz, D.~Jamiola, J.~Choma, and M.~Jaroniec, \enquote{{Silica-metal
  core-shell nanostructures},} {\protect\JournalTitle{Advances in Colloid and
  Interface Science}} \textbf{170}, 28--47 (2012).

\bibitem{Montano-Priede2017}
J.~L. Monta{\~{n}}o-Priede, O.~Pe{\~{n}}a-Rodr{\'{i}}guez, and U.~Pal,
  \enquote{{Near-electric-field tuned plasmonic Au@SiO2 and Ag@SiO2
  nanoparticles for efficient utilization in luminescence enhancement and
  surface-enhanced spectroscopy},} {\protect\JournalTitle{Journal of Physical
  Chemistry C}} \textbf{121}, 23062--23071 (2017).

\bibitem{Montano-Priede2017a}
J.~L. Monta{\~{n}}o-Priede, J.~P. Coelho, A.~Guerrero-Mart{\'{i}}nez,
  O.~Pe{\~{n}}a-Rodr{\'{i}}guez, and U.~Pal, \enquote{{Fabrication of
  monodispersed Au@SiO2 nanoparticles with highly stable silica layers by
  ultrasound-assisted St{\"{o}}ber method},} {\protect\JournalTitle{Journal of
  Physical Chemistry C}} \textbf{121}, 9543--9551 (2017).

\bibitem{Wang2018h}
P.~Wang, A.~V. Krasavin, F.~N. Viscomi, A.~M. Adawi, J.-S.~G. Bouillard,
  L.~Zhang, D.~J. Roth, L.~Tong, and A.~V. Zayats, \enquote{{Metaparticles:
  dressing nano-objects with a hyperbolic coating},}
  {\protect\JournalTitle{Laser {\&} Photonics Reviews}} \textbf{12}, 1800179
  (2018).

\bibitem{Aden1951}
A.~L. Aden and M.~Kerker, \enquote{{Scattering of electromagnetic waves from
  two concentric spheres},} {\protect\JournalTitle{Journal of Applied Physics}}
  \textbf{22}, 1242--1246 (1951).

\bibitem{Kaiser1994}
T.~Kaiser, S.~Lange, and G.~Schweiger, \enquote{{Structural resonances in a
  coated sphere: investigation of the volume-averaged source function and
  resonance positions},} {\protect\JournalTitle{Applied Optics}} \textbf{33},
  7789 (1994).

\bibitem{Lock1994}
J.~A. Lock, J.~M. Jamison, and C.-Y. Lin, \enquote{{Rainbow scattering by a
  coated sphere},} {\protect\JournalTitle{Applied Optics}} \textbf{33}, 4677
  (1994).

\bibitem{Sinzig1994}
J.~Sinzig and M.~Quinten, \enquote{{Scattering and absorption by spherical
  multilayer particles},} {\protect\JournalTitle{Applied Physics A Solids and
  Surfaces}} \textbf{58}, 157--162 (1994).

\bibitem{Bhandari1985}
R.~Bhandari, \enquote{{Scattering coefficients for a multilayered sphere:
  analytic expressions and algorithms},} {\protect\JournalTitle{Applied
  Optics}} \textbf{24}, 1960--1967 (1985).

\bibitem{Mackowski1990}
D.~W. Mackowski, R.~A. Altenkirch, and M.~P. Menguc, \enquote{{Internal
  absorption cross sections in a stratified sphere},}
  {\protect\JournalTitle{Applied Optics}} \textbf{29}, 1551--1559 (1990).

\bibitem{Li2007}
R.~Li, X.~Han, L.~Shi, K.~F. Ren, and H.~Jiang, \enquote{{Debye series for
  Gaussian beam scattering by a multilayered sphere},}
  {\protect\JournalTitle{Applied Optics}} \textbf{46}, 4804--4812 (2007).

\bibitem{Wang2011}
J.~J. Wang, G.~Gouesbet, G.~Gr{\'{e}}han, Y.~P. Han, and S.~Saengkaew,
  \enquote{{Morphology-dependent resonances in an eccentrically layered sphere
  illuminated by a tightly focused off-axis Gaussian beam: parallel and
  perpendicular beam incidence},} {\protect\JournalTitle{Journal of the Optical
  Society of America A}} \textbf{28}, 1849--1859 (2011).

\bibitem{Onofri1995}
F.~Onofri, G.~Gr{\'{e}}han, and G.~Gouesbet, \enquote{{Electromagnetic
  scattering from a multilayered sphere located in an arbitrary beam},}
  {\protect\JournalTitle{Applied Optics}} \textbf{34}, 7113--7124 (1995).

\bibitem{Wu1997a}
Z.~S. Wu, L.~X. Guo, K.~F. Ren, G.~Gouesbet, and G.~Gr{\'{e}}han,
  \enquote{{Improved algorithm for electromagnetic scattering of plane waves
  and shaped beams by multilayered spheres},} {\protect\JournalTitle{Applied
  Optics}} \textbf{36}, 5188--5198 (1997).

\bibitem{Moroz2005}
A.~Moroz, \enquote{{A recursive transfer-matrix solution for a dipole radiating
  inside and outside a stratified sphere},} {\protect\JournalTitle{Ann. Phys.
  (NY)}} \textbf{315}, 352--418 (2005).

\bibitem{Moroz2005a}
A.~Moroz, \enquote{{Spectroscopic properties of a two-level atom interacting
  with a complex spherical nanoshell},} {\protect\JournalTitle{Chemical
  Physics}} \textbf{317}, 1--15 (2005).

\bibitem{Rasskazov19JOSAA}
I.~L. Rasskazov, A.~Moroz, and P.~S. Carney, \enquote{{Electromagnetic energy
  in multilayered spherical particles},} {\protect\JournalTitle{Journal of the
  Optical Society of America A}} \textbf{36}, 1591--1601 (2019).

\bibitem{Schelm2005a}
S.~Schelm and G.~B. Smith, \enquote{{Internal electric field densities of metal
  nanoshells},} {\protect\JournalTitle{Journal of Physical Chemistry B}}
  \textbf{109}, 1689--1694 (2005).

\bibitem{Toon1981}
O.~B. Toon and T.~P. Ackerman, \enquote{{Algorithms for the calculation of
  scattering by stratified spheres},} {\protect\JournalTitle{Applied Optics}}
  \textbf{20}, 3657--3660 (1981).

\bibitem{Wu1991}
Z.~S. Wu and Y.~P. Wang, \enquote{{Electromagnetic scattering for multilayered
  sphere: recursive algorithms},} {\protect\JournalTitle{Radio Science}}
  \textbf{26}, 1393--1401 (1991).

\bibitem{Yang2003}
W.~Yang, \enquote{{Improved recursive algorithm for light scattering by a
  multilayered sphere},} {\protect\JournalTitle{Applied Optics}} \textbf{42},
  1710--1720 (2003).

\bibitem{Majic2020}
M.~Majic and E.~C. Le~Ru, \enquote{{Numerically stable formulation of Mie
  theory for an emitter close to a sphere},} {\protect\JournalTitle{Applied
  Optics}} \textbf{59}, 1293--1300 (2020).

\bibitem{wvsccodes}
A.~Moroz, \enquote{{http://wave-scattering.com/codes.html},} .

\bibitem{Bohren1998}
C.~F. Bohren and D.~R. Huffman, \emph{{Absorption and scattering of light by
  small particles}} (Wiley-VCH Verlag GmbH, Weinheim, Germany, 1998).

\bibitem{Ladutenko2017}
K.~Ladutenko, U.~Pal, A.~Rivera, and O.~Pe{\~{n}}a-Rodr{\'{i}}guez,
  \enquote{{Mie calculation of electromagnetic near-field for a multilayered
  sphere},} {\protect\JournalTitle{Computer Physics Communications}}
  \textbf{214}, 225--230 (2017).

\bibitem{Abeles1948}
F.~Abel{\`{e}}s, \enquote{{Sur la propagation des ondes
  {\'{e}}lectromagn{\'{e}}tiques dans les milieux sratifi{\'{e}}s},}
  {\protect\JournalTitle{Annales de Physique}} \textbf{12}, 504--520 (1948).

\bibitem{Abeles1950}
F.~Abel{\`{e}}s, \enquote{{Recherches sur la propagation des ondes
  {\'{e}}lectromagn{\'{e}}tiques sinuso{\"{i}}dales dans les milieux
  stratifi{\'{e}}s},} {\protect\JournalTitle{Annales de Physique}} \textbf{12},
  596--640 (1950).

\bibitem{Abeles1950a}
F.~Abel{\`{e}}s, \enquote{{Recherches sur la propagation des ondes
  {\'{e}}lectromagn{\'{e}}tiques sinuso{\"{i}}dales dans les milieux
  stratifi{\'{e}}s},} {\protect\JournalTitle{Annales de Physique}} \textbf{12},
  706--782 (1950).

\bibitem{Born2013}
M.~Born and E.~Wolf, \emph{{Principles of optics: Electromagnetic theory of
  propagation, interference and diffraction of light}} (Elsevier, 2013), 6th
  ed.

\bibitem{GarciadeAbajo2002}
F.~J. Garc{\'{i}}a~de Abajo and A.~Howie, \enquote{{Retarded field calculation
  of electron energy loss in inhomogeneous dielectrics},}
  {\protect\JournalTitle{Physical Review B}} \textbf{65}, 115418 (2002).

\bibitem{Hohenester2012}
U.~Hohenester and A.~Tr{\"{u}}gler, \enquote{{MNPBEM - A Matlab toolbox for the
  simulation of plasmonic nanoparticles},} {\protect\JournalTitle{Computer
  Physics Communications}} \textbf{183}, 370--381 (2012).

\bibitem{Hohenester2014}
U.~Hohenester, \enquote{{Simulating electron energy loss spectroscopy with the
  MNPBEM toolbox},} {\protect\JournalTitle{Computer Physics Communications}}
  \textbf{185}, 1177--1187 (2014).

\bibitem{Waxenegger2015}
J.~Waxenegger, A.~Tr{\"{u}}gler, and U.~Hohenester, \enquote{{Plasmonics
  simulations with the MNPBEM toolbox: Consideration of substrates and layer
  structures},} {\protect\JournalTitle{Computer Physics Communications}}
  \textbf{193}, 138--150 (2015).

\bibitem{Hohenester2018}
U.~Hohenester, \enquote{{Making simulations with the MNPBEM toolbox big:
  Hierarchical matrices and iterative solvers},}
  {\protect\JournalTitle{Computer Physics Communications}} \textbf{222},
  209--228 (2018).

\bibitem{Jackson1999}
J.~D. Jackson, \emph{{Classical electrodynamics}} (John Wiley {\&} Sons, Inc.,
  1999), 3rd ed.

\bibitem{Kerker1980}
M.~Kerker, D.-S. Wang, and H.~Chew, \enquote{{Surface enhanced Raman scattering
  (SERS) by molecules adsorbed at spherical particles},}
  {\protect\JournalTitle{Applied Optics}} \textbf{19}, 3373--3388 (1980).

\bibitem{Mishchenko2002}
M.~I. Mishchenko, L.~D. Travis, and A.~A. Lacis, \emph{{Scattering, Absorption,
  and Emission of Light by Small Particles}} (Cambridge University Press,
  Cambridge, UK, 2002).

\bibitem{Abramowitz1973}
M.~Abramowitz and I.~A. Stegun, \emph{{Handbook of Mathematical Functions}}
  (Dover Publications, New York, 1973).

\bibitem{Bott1987}
A.~Bott and W.~Zdunkowski, \enquote{{Electromagnetic energy within dielectric
  spheres},} {\protect\JournalTitle{Journal of the Optical Society of America
  A}} \textbf{4}, 1361--1365 (1987).

\bibitem{Loudon1970}
R.~Loudon, \enquote{{The propagation of electromagnetic energy through an
  absorbing dielectric},} {\protect\JournalTitle{Journal of Physics A: General
  Physics}} \textbf{3}, 233--245 (1970).

\bibitem{Blaber2009}
M.~G. Blaber, M.~D. Arnold, and M.~J. Ford, \enquote{{Search for the ideal
  plasmonic nanoshell: the effects of surface scattering and alternatives to
  gold and silver},} {\protect\JournalTitle{Journal of Physical Chemistry C}}
  \textbf{113}, 3041--3045 (2009).

\bibitem{Ordal1985}
M.~A. Ordal, R.~J. Bell, R.~W. Alexander, L.~L. Long, and M.~R. Querry,
  \enquote{{Optical properties of fourteen metals in the infrared and far
  infrared: Al, Co, Cu, Au, Fe, Pb, Mo, Ni, Pd, Pt, Ag, Ti, V, and W},}
  {\protect\JournalTitle{Applied Optics}} \textbf{24}, 4493--4499 (1985).

\bibitem{Arruda2010}
T.~J. Arruda and A.~S. Martinez, \enquote{{Electromagnetic energy within
  magnetic spheres},} {\protect\JournalTitle{Journal of the Optical Society of
  America A}} \textbf{27}, 992--1001 (2010).

\bibitem{Arruda2012}
T.~J. Arruda, F.~A. Pinheiro, and A.~S. Martinez, \enquote{{Electromagnetic
  energy within coated spheres containing dispersive metamaterials},}
  {\protect\JournalTitle{Journal of Optics}} \textbf{14}, 065101 (2012).

\bibitem{Wiscombe1980}
W.~J. Wiscombe, \enquote{{Improved Mie scattering algorithms},}
  {\protect\JournalTitle{Applied Optics}} \textbf{19}, 1505--1509 (1980).

\bibitem{Allardice2014}
J.~R. Allardice and E.~C. Le~Ru, \enquote{{Convergence of Mie theory series:
  criteria for far-field and near-field properties},}
  {\protect\JournalTitle{Applied Optics}} \textbf{53}, 7224--7229 (2014).

\bibitem{Moroz2010}
A.~Moroz, \enquote{{Non-radiative decay of a dipole emitter close to a metallic
  nanoparticle: Importance of higher-order multipole contributions},}
  {\protect\JournalTitle{Optics Communications}} \textbf{283}, 2277--2287
  (2010).

\bibitem{Moroz2008}
A.~Moroz, \enquote{{Electron mean free path in a spherical shell geometry},}
  {\protect\JournalTitle{Journal of Physical Chemistry C}} \textbf{112},
  10641--10652 (2008).

\bibitem{Averitt1999}
R.~D. Averitt, S.~L. Westcott, and N.~J. Halas, \enquote{{Linear optical
  properties of gold nanoshells},} {\protect\JournalTitle{Journal of the
  Optical Society of America B}} \textbf{16}, 1824--1832 (1999).

\bibitem{Ruppin2015}
R.~Ruppin, \enquote{{Nanoshells with a gain layer: the effects of surface
  scattering},} {\protect\JournalTitle{Journal of Optics}} \textbf{17}, 125004
  (2015).

\bibitem{Zakomirnyi20PCCP}
V.~I. Zakomirnyi, I.~L. Rasskazov, L.~K. S{\o}rensen, P.~S. Carney,
  Z.~Rinkevicius, and H.~{\AA}gren, \enquote{{Plasmonic nano-shells: atomistic
  discrete interaction versus classic electrodynamics models},}
  {\protect\JournalTitle{Physical Chemistry Chemical Physics}}  (2020).

\bibitem{Meng2017}
L.~Meng, R.~Yu, M.~Qiu, and F.~J. Garc{\'{i}}a~de Abajo, \enquote{{Plasmonic
  nano-oven by concatenation of multishell photothermal enhancement},}
  {\protect\JournalTitle{ACS Nano}} \textbf{11}, 7915--7924 (2017).

\bibitem{Kodali2010}
A.~K. Kodali, M.~V. Schulmerich, R.~Palekar, X.~Llora, and R.~Bhargava,
  \enquote{{Optimized nanospherical layered alternating metal-dielectric probes
  for optical sensing},} {\protect\JournalTitle{Optics Express}} \textbf{18},
  23302 (2010).

\bibitem{Kodali2010a}
A.~K. Kodali, X.~Llora, and R.~Bhargava, \enquote{{Optimally designed
  nanolayered metal-dielectric particles as probes for massively multiplexed
  and ultrasensitive molecular assays},} {\protect\JournalTitle{Proceedings of
  the National Academy of Sciences}} \textbf{107}, 13620--13625 (2010).

\bibitem{Khlebtsov2017}
N.~G. Khlebtsov and B.~N. Khlebtsov, \enquote{{Optimal design of gold
  nanomatryoshkas with embedded Raman reporters},}
  {\protect\JournalTitle{Journal of Quantitative Spectroscopy and Radiative
  Transfer}} \textbf{190}, 89--102 (2017).

\bibitem{Lee2018}
J.~Y. Lee, A.~E. Miroshnichenko, and R.-K. Lee, \enquote{{Simultaneously nearly
  zero forward and nearly zero backward scattering objects},}
  {\protect\JournalTitle{Optics Express}} \textbf{26}, 30393--30399 (2018).

\bibitem{Ruan2010}
Z.~Ruan and S.~Fan, \enquote{{Superscattering of light from subwavelength
  nanostructures},} {\protect\JournalTitle{Physical Review Letters}}
  \textbf{105}, 013901 (2010).

\bibitem{Ruan2011}
Z.~Ruan and S.~Fan, \enquote{{Design of subwavelength superscattering
  nanospheres},} {\protect\JournalTitle{Applied Physics Letters}} \textbf{98},
  043101 (2011).

\bibitem{Argyropoulos2013}
C.~Argyropoulos, F.~Monticone, G.~D'Aguanno, and A.~Al{\`{u}},
  \enquote{{Plasmonic nanoparticles and metasurfaces to realize Fano spectra at
  ultraviolet wavelengths},} {\protect\JournalTitle{Applied Physics Letters}}
  \textbf{103}, 143113 (2013).

\bibitem{Fleury2014}
R.~Fleury, J.~Soric, and A.~Al{\`{u}}, \enquote{{Physical bounds on absorption
  and scattering for cloaked sensors},} {\protect\JournalTitle{Physical Review
  B}} \textbf{89}, 045122 (2014).

\bibitem{Lepeshov2019}
S.~Lepeshov, A.~Krasnok, and A.~Al{\`{u}},
  \enquote{{Nonscattering-to-superscattering switch with phase-change
  materials},} {\protect\JournalTitle{ACS Photonics}} \textbf{6}, 2126--2132
  (2019).

\bibitem{Yezekyan2020}
T.~Yezekyan, K.~V. Nerkararyan, and S.~I. Bozhevolnyi, \enquote{{Maximizing
  absorption and scattering by spherical nanoparticles},}
  {\protect\JournalTitle{Optics Letters}} \textbf{45}, 1531--1534 (2020).

\bibitem{Tuersun2013}
P.~Tuersun and X.~Han, \enquote{{Optical absorption analysis and optimization
  of gold nanoshells},} {\protect\JournalTitle{Applied Optics}} \textbf{52},
  1325 (2013).

\bibitem{Ladutenko2015}
K.~Ladutenko, P.~Belov, O.~Pe{\~{n}}a-Rodr{\'{i}}guez, A.~Mirzaei, A.~E.
  Miroshnichenko, and I.~V. Shadrivov, \enquote{{Superabsorption of light by
  nanoparticles},} {\protect\JournalTitle{Nanoscale}} \textbf{7}, 18897--18901
  (2015).

\bibitem{Xue2016}
X.~Xue, V.~Sukhotskiy, and E.~P. Furlani, \enquote{{Optimization of optical
  absorption of colloids of SiO2@Au and Fe3O4@Au nanoparticles with
  constraints},} {\protect\JournalTitle{Scientific Reports}} \textbf{6}, 35911
  (2016).

\bibitem{Monticone2014}
F.~Monticone and A.~Al{\`{u}}, \enquote{{Embedded photonic eigenvalues in 3D
  nanostructures},} {\protect\JournalTitle{Physical Review Letters}}
  \textbf{112}, 213903 (2014).

\bibitem{Gordon2007}
J.~A. Gordon and R.~W. Ziolkowski, \enquote{{The design and simulated
  performance of a coated nano-particle laser},} {\protect\JournalTitle{Optics
  Express}} \textbf{15}, 2622--2653 (2007).

\bibitem{Pezzi2019}
L.~Pezzi, M.~A. Iat{\`{i}}, R.~Saija, A.~De~Luca, and O.~M. Marag{\`{o}},
  \enquote{{Resonant coupling and gain singularities in metal/dielectric
  multishells: Quasi-static versus T-matrix calculations},}
  {\protect\JournalTitle{Journal of Physical Chemistry C}} \textbf{123},
  29291--29297 (2019).

\bibitem{Miroshnichenko2010a}
A.~E. Miroshnichenko, \enquote{{Off-resonance field enhancement by spherical
  nanoshells},} {\protect\JournalTitle{Physical Review A}} \textbf{81}, 053818
  (2010).

\bibitem{Melnyk1970}
A.~R. Melnyk and M.~J. Harrison, \enquote{{Theory of optical excitation of
  plasmons in metals},} {\protect\JournalTitle{Physical Review B}} \textbf{2},
  835--850 (1970).

\bibitem{Anderegg1971}
M.~Anderegg, B.~Feuerbacher, and B.~Fitton, \enquote{{Optically excited
  longitudinal plasmons in potassium},} {\protect\JournalTitle{Physical Review
  Letters}} \textbf{27}, 1565--1568 (1971).

\bibitem{Ruppin1975}
R.~Ruppin, \enquote{{Optical properties of small metal spheres},}
  {\protect\JournalTitle{Physical Review B}} \textbf{11}, 2871--2876 (1975).

\bibitem{Leung1990}
P.~T. Leung, \enquote{{Decay of molecules at spherical surfaces: Nonlocal
  effects},} {\protect\JournalTitle{Physical Review B}} \textbf{42}, 7622--7625
  (1990).

\bibitem{Rojas1988}
R.~Rojas, F.~Claro, and R.~Fuchs, \enquote{{Nonlocal response of a small coated
  sphere},} {\protect\JournalTitle{Physical Review B}} \textbf{37}, 6799--6807
  (1988).

\bibitem{David2011}
C.~David and F.~J. Garc{\'{i}}a~de Abajo, \enquote{{Spatial nonlocality in the
  optical response of metal nanoparticles},} {\protect\JournalTitle{Journal of
  Physical Chemistry C}} \textbf{115}, 19470--19475 (2011).

\bibitem{Huang2014}
Y.~Huang and L.~Gao, \enquote{{Superscattering of light from core-shell
  nonlocal plasmonic nanoparticles},} {\protect\JournalTitle{Journal of
  Physical Chemistry C}} \textbf{118}, 30170--30178 (2014).

\bibitem{Mortensen2014}
N.~A. Mortensen, S.~Raza, M.~Wubs, T.~S{\o}ndergaard, and S.~I. Bozhevolnyi,
  \enquote{{A generalized non-local optical response theory for plasmonic
  nanostructures},} {\protect\JournalTitle{Nature Communications}} \textbf{5},
  3809 (2014).

\bibitem{Dong2017c}
T.~Dong, Y.~Shi, H.~Liu, F.~Chen, X.~Ma, and R.~Mittra, \enquote{{Investigation
  on plasmonic responses in multilayered nanospheres including asymmetry and
  spatial nonlocal effects},} {\protect\JournalTitle{Journal of Physics D:
  Applied Physics}} \textbf{50}, 495302 (2017).

\bibitem{Eremin2019}
Y.~Eremin, A.~Doicu, and T.~Wriedt, \enquote{{Extension of the discrete sources
  method to investigate the non-local effect influence on non-spherical
  core-shell particles},} {\protect\JournalTitle{Journal of Quantitative
  Spectroscopy and Radiative Transfer}} \textbf{235}, 300--308 (2019).

\bibitem{Bohren1974}
C.~F. Bohren, \enquote{{Light scattering by an optically active sphere},}
  {\protect\JournalTitle{Chemical Physics Letters}} \textbf{29}, 458--462
  (1974).

\bibitem{Bohren1975}
C.~F. Bohren, \enquote{{Scattering of electromagnetic waves by an optically
  active spherical shell},} {\protect\JournalTitle{Journal of Chemical
  Physics}} \textbf{62}, 1566--1571 (1975).

\bibitem{Lakhtakia1985}
A.~Lakhtakia, V.~K. Varadan, and V.~V. Varadan, \enquote{{Scattering and
  absorption characteristics of lossy dielectric, chiral, nonspherical
  objects},} {\protect\JournalTitle{Applied Optics}} \textbf{24}, 4146--4154
  (1985).

\bibitem{Engheta1990}
N.~Engheta and M.~W. Kowarz, \enquote{{Antenna radiation in the presence of a
  chiral sphere},} {\protect\JournalTitle{Journal of Applied Physics}}
  \textbf{67}, 639--647 (1990).

\bibitem{Yokota2001}
M.~Yokota, S.~He, and T.~Takenaka, \enquote{{Scattering of a Hermite-Gaussian
  beam field by a chiral sphere},} {\protect\JournalTitle{Journal of the
  Optical Society of America A}} \textbf{18}, 1681--1689 (2001).

\bibitem{Guzatov2012a}
D.~V. Guzatov and V.~V. Klimov, \enquote{{The influence of chiral spherical
  particles on the radiation of optically active molecules},}
  {\protect\JournalTitle{New Journal of Physics}} \textbf{14}, 123009 (2012).

\bibitem{Klimov2012}
V.~V. Klimov, D.~V. Guzatov, and M.~Ducloy, \enquote{{Engineering of radiation
  of optically active molecules with chiral nano-meta-particles},}
  {\protect\JournalTitle{EPL}} \textbf{97}, 47004 (2012).

\bibitem{Arruda2013a}
T.~J. Arruda, F.~A. Pinheiro, and A.~S. Martinez, \enquote{{Electromagnetic
  energy within single-resonance chiral metamaterial spheres},}
  {\protect\JournalTitle{Journal of the Optical Society of America A}}
  \textbf{30}, 1205--1212 (2013).

\bibitem{Gouesbet1988}
G.~Gouesbet, B.~Maheu, and G.~Gr{\'{e}}han, \enquote{{Light scattering from a
  sphere arbitrarily located in a Gaussian beam, using a Bromwich
  formulation},} {\protect\JournalTitle{Journal of the Optical Society of
  America A}} \textbf{5}, 1427--1443 (1988).

\bibitem{Mojarad2009}
N.~M. Mojarad, G.~Zumofen, V.~Sandoghdar, and M.~Agio, \enquote{{Metal
  nanoparticles in strongly confined beams: transmission, reflection and
  absorption},} {\protect\JournalTitle{Journal of the European Optical Society:
  Rapid Publications}} \textbf{4}, 09014 (2009).

\bibitem{DeDood2001}
M.~J.~A. de~Dood, L.~H. Slooff, A.~Polman, A.~Moroz, and A.~van Blaaderen,
  \enquote{{Modified spontaneous emission in erbium-doped SiO2 spherical
  colloids},} {\protect\JournalTitle{Applied Physics Letters}} \textbf{79},
  3585--3587 (2001).

\bibitem{DeDood2001a}
M.~J.~A. de~Dood, L.~H. Slooff, A.~Polman, A.~Moroz, and A.~van Blaaderen,
  \enquote{{Local optical density of states in SiO2 spherical microcavities:
  Theory and experiment},} {\protect\JournalTitle{Physical Review A}}
  \textbf{64}, 033807 (2001).

\bibitem{Wiecha2018}
P.~R. Wiecha, A.~Arbouet, A.~Cuche, V.~Paillard, and C.~Girard, \enquote{{Decay
  rate of magnetic dipoles near nonmagnetic nanostructures},}
  {\protect\JournalTitle{Physical Review B}} \textbf{97}, 085411 (2018).

\end{thebibliography}

\end{document}